\documentclass[hyper(*)]{JHEP3} 

\usepackage{epsfig,multicol,bbm}

\newcommand\fverb{\setbox\fverbbox=\hbox\bgroup\verb}
\newcommand\fverbdo{\egroup\medskip\noindent%
            \fbox{\unhbox\fverbbox}\ }
\newcommand\fverbit{\egroup\item[\fbox{\unhbox\fverbbox}]}
\newbox\fverbbox

\title{More Three Dimensional Mirror Pairs}

\author{Dimitri Nanopoulos $^{1,2,3,a}$, and Dan Xie $^{1,b}$

\\$^{1}$George P. and Cynthia W.Mitchell Institute for Fundamental Physics,
Texas A\&M University, College Station, TX 77843,USA \\
$^{2}$Astroparticle physics Group, Houston Advanced Research
Center (HARC), Mitchell Campus, Woodlands, TX 77381, USA\\
$^{3}$Academy of Athens, Division of Nature Sciences, 28
panepistimiou Avenue, Athens 10679, Greece\\
E-mail: \email{$^{a}$dimitri@physics.tamu.edu,
$^{b}$fogman@neo.tamu.edu}}

\preprint{MIFPA-10-51}  

\abstract{We found a lot of new three dimensional ${\cal N}=4$ mirror pairs
generalizing previous considerations on three dimensional generalized quiver gauge theories. We recovered
almost all previous discovered mirror pairs with these constructions. One side of these mirror
pairs  are always the conventional quiver gauge theories.
One of our result can also be used to determine the matter content and weakly coupled gauge groups
of four dimensional ${\cal N}=2$ generalized
quiver gauge theories derived from six dimensional $A_N$ and $D_N$ theory, therefore we explicitly
constructed four dimensional S-duality pairs. }


\dedicated{}

\begin{document}


\section{Introduction}
Last year, Gaiotto \cite{Gaiotto1} found a large class of four dimensional (4d) ${\cal N}=2$
superconformal field theories (SCFT) motivated by earlier remarkable observation
on S-duality of certain ${\cal N}=2$ SCFT \cite{Argy}. These theories are derived
by compactifying six dimensional $(0,2)$ $A_{N-1}$ theory on a Riemann surface
with punctures. The S-duality group is realized as the modular group of
the Riemann surface. Different weakly coupled duality frames correspond
to different degeneration limits of this Riemann surface. The
weakly coupled gauge groups in each duality frame can be
derived from the information encoded in the punctures. In general, these
generalized quiver gauge theories do not have conventional lagrangian
descriptions in any duality frame since the matter systems are usually strongly coupled.

It would be very helpful if we can find a lagrangian description in
some sense for those strongly coupled theories. Compactifying these
theories down to three dimension (3d) seems to be a good idea \cite{3d}.
If we further compactify the theory on a circle and flow to IR limit,
we will get a fixed point with three dimensional ${\cal N}=4$ supersymmetry (We call
these theories $A$). Three dimensional ${\cal N}=4$ supersymmetric field theory has
$SU(2)_L\times SU(2)_R$ R symmetry. The moduli space of vacua of 3d theory has
Coulomb branch and Higgs branch. There is an interesting mirror symmetry for some of 3d ${\cal N}=4$
theories discovered in middle nighties by Intriligator and Seiberg \cite{Mr}. Mirror symmetry states
that given theory $A$, there is another theory $B$ for which the Higgs branch is the Coulomb branch of
$A$ and vice versa (We usually write the UV theory for $B$ and simply call it $B$, to find the mirror, we
mean to find the UV theory $B$). The mirror theory does not necessarily have a
lagrangian description, for instance, the quiver gauge theory with
$E$ type dynkin diagram shape do not have the lagrangian description
as first discovered in \cite{Mr}.

Surprisingly, it was found in \cite{BYX} that all the generalized
quiver gauge theory $A$ have the 3d mirror theory $B$ which has the conventional lagrangian
description: it is just a star-shaped conventional quiver gauge theory.
With this mirror, one can learn a lot about the four dimensional
theory, for instance, one can figure out the full flavor symmetry
of the theory $A$ using the monopole operators of the mirror $B$, etc.

A purely field theory derivation for this result is also presented
in \cite{BYX}. The result is nicely derived by representing three dimensional theory $A$ as compactifying
 four dimensional ${\cal N}=4$ Supersymmetric Yang-Mills (SYM) on a one dimensional graph. The graph
is just the dual graph of the punctured Riemann surface
with the boundary conditions on the external leg
labeled by the same Young tableaux needed for four dimensional theory.
The internal leg represents the weakly coupled gauge groups. The mirror symmetry is just
the S-duality of ${\cal N}=4$ SYM on the graph.
${\cal N}=4$ SYM on the half space is extensively studied in \cite{GW1, GW2}.
It turns out that these constructions are very useful to derive new mirror pairs.

In this paper, we will generalize the construction in \cite{BYX} and
find a lot of new mirror pairs. With these new construction, we almost
recover all the previously discovered examples of mirror pairs \cite{{Ooguri1,
Ooguri2, Feng, HW, Kapustin1, Kapustin2,Vafa}}.

The theory $A$ considered in \cite{BYX} is conformal in four dimension.
The first generalization is to add fundamentals to each weakly coupled
gauge group to form theory $\tilde{A}$(in this paper, we use $\tilde{A}$
to refer the new theory based on $A$, and likewise use $\tilde{B}$ for the mirror of $\tilde{A}$).
It is natural to consider those theories in three dimensions though in four
dimension those theories are not asymptotical free.

We add ``D5'' branes on the internal leg of the graph representation of $A$
and find the mirror of the graph
by simply do S-duality and turn the ``D5'' branes to ``NS5'' brane. The problem
is that the IR limit of the graph representation is not the same as $\tilde{A}$,
we find a way to extract the mirror theory $\tilde{B}$ from the graph mirror.
The mirrors are different for different weakly coupled descriptions of the
generalized quiver gauge theory, this is in contrast with the conformal case for which
the mirror is independent of the duality frame.

The mirror theory $\tilde{B}$ can be used to probe theory $A$, since $\tilde{B}$ depends
on the duality frames of 4d theory. For
example, we can read the weakly coupled gauge groups in
different duality frames by counting the change of Coulomb branch dimension of $\tilde{B}$
with respect to $B$.
The number of fundamentals on the gauge group can also be determined by
finding out the change of the symmetry on the Coulomb branch of $\tilde{B}$ with respect to $B$.
The results for $A_{N-1}$ theory
are in agreement with the other methods \cite{dan1}. This is particularly useful
for the theory derived from 6d $D_N$ theory for which no method determining
the weakly coupled gauge groups has found yet. One can also completely determine the
matter contents in each duality frame.

The second generalization is to gauge some of the $U(1)$ flavor symmetries of the
theory $A$ to get theory $\tilde{A}$, the mirror can be found by counting the
change of the Higgs and Coulomb branch dimensions: we only need to ungauge the
$U(1)$ gauge groups in the mirror.

The above mirror theory has the quiver tail without any fundamentals. The third generalization
is to change the boundary condition of the graph so the mirror has a general quiver tail. A large class
of boundary 3d SCFT and its mirror are found in \cite{GW2}, we can use their
results to construct mirror pairs $\tilde{A}$ and $\tilde{B}$ so that $\tilde{B}$ has more
general quiver tails.

Finally, we consider four dimensional theory derived from putting irregular
singularities on the Riemann sphere \cite{dan2}. The mirror of the three dimensional cousin
is quite interesting and is very useful for studying the four dimensional theory.
The mirror is also a quiver gauge theory but with more exotic shapes and more
than one bi-fundamentals connecting the quiver nodes.

This paper is organized as follows: In section 2, we summarize the results
in \cite{BYX}; In section 3, we add more fundamentals on
theory $A$ reviewed in section 2, the unitary quiver and orthosymplectic quiver (quiver with alternative
orthogonal and symplectic gauge groups) are both discussed, we also discuss how to use this result to
probe four dimensional generalized quiver gauge theory; Section 4 discusses how to find the mirror if we
gauge the $U(1)$ flavor symmetry of theory $A$; In section 5, the boundary conditions
are changed so that we can have general quiver tails in the mirror; In section 6,
we consider four dimensional theory derived from 6d theory on a Riemann sphere
with irregular singularities, the mirror theory for 3d cousin is
also a quiver gauge theory but with exotic shape. We conclude in section 7 by giving
some further directions.

\section{Review}
Three dimensional ${\cal N}=4$ gauge theory has $SU(2)_L\times SU(2)_R$ R symmetry. This
can be seen from the compactification of 6d  ${\cal N}=1$ theory: $SU(2)_R$
is the R symmetry of the 6d theory while $SU(2)_L$ symmetry comes from the
rotation group of the three dimensional space on which we do the reduction. The moduli space
of the vacua has Coulomb branch and Higgs branch (we also have the mixed branch).
The Higgs branch is a Hyperkahler manifold whose Kahler form transforms under $SU(2)_R$ and
invariant under the $SU(2)_L$. There usually are global symmetries acting on Higgs branch, when we have
a lagrangian description, the global symmetry can be read readily, we can turn on mass terms and
preserve ${\cal N}=4$ supersymmetry. The Coulomb branch
is also a Hyperkahler manifold whose kahler form transforms under $SU(2)_L$ and
invariant under $SU(2)_R$. Usually there is only a $U(1)$ global symmetry arising from the shift
symmetry of the photon, but sometimes the symmetry is enhanced due to monopole
operators \cite{Mon,GW2}; if there are $U(1)$ factors in gauge group, we can turn on Fayet-lliopoulos (FI) terms and
preserve the same number of supersymmetry.
For some theories, the Higgs branch and Coulomb branch intersects at a single
point, and there is an interacting SCFT on which both $SU(2)_L$ and $SU(2)_R$ acts. This SCFT
is the IR fixed point under the RG flow of the theory.

Suppose we have two three dimensional ${\cal N}=4$ theory $A$ and $B$, and both theories flow to
non-trivial IR fixed points ${\cal A}$ and ${\cal B}$. We say they are mirror pairs if the Higgs
branch of ${\cal A}$ is identical to the Coulomb branch of ${\cal B}$ and vice versa \cite{Mr}.
The mass terms are identified with the FI terms under mirror symmetry.
Since Coulomb branch gets quantum corrections and Higgs branch has the non-renormalization property and
is exact by doing classical calculation, the quantum effects of one theory is
captured by classical effects of another theory. The IR fixed points are usually strongly
coupled, we mainly use the UV theory $A$ and $B$ to learn their IR behavior and simply
states the theory $A$ and $B$ are mirror pairs.

In \cite{BYX}, a large class of mirror pairs are found. Theory $A$ arises from compactifying
4d ${\cal N}=2$ SCFT theory found in \cite{Gaiotto1} on a circle; Theory $B$ is a star-shaped quiver.
Four dimensional theory is realized as compactifying six dimensional $(0,2)$ theory on
a Riemann surface $\Sigma$ with punctures which are classified by Young tableaux \cite{Gaiotto1,dan3}.
The Hitchin's equation defined on Riemann surface is the BPS equation and whose moduli space
with specified boundary condition at the puncture is the Coulomb branch of the three dimensional theory.
The boundary condition of the Hitchin's equation is a regular singularity for this class of theories.
The Hitchin's moduli space can be approximately by a quiver as discovered by Boalch \cite{Bol}, it turns
out that this quiver is the mirror quiver for the theory $A$. We only consider those theories
for which the Hitchin's system is irreducible \cite{dan1}. In physics language, this means that the quiver gauge
theory has a dimension $N$ operators in the Coulomb branch.

In general, we can not write a lagrangian description for the theory; The weakly coupled
gauge group and flavor symmetries can be determined using the information on the puncture \cite{Gaiotto1, dan1}
, we also know the flavor symmetry. These theories and its generalization are further studied in \cite{SO,web,N1,dan4,yuji2,stony, stony1,dist,mal}.
Various $S$-duality frames of 4d theory are identified with the different degeneration limits of the punctured
Riemann surface.

We further compactify theory $A$ on a circle $S$ and get a 3d ${\cal N}=4$ theory. The compact space is
$\Sigma\times S$.
We can model each leg in pants decomposition of the punctured Riemann surface  as
a cylinder $S^1\times I$, then the three dimensional space
we do the reduction on this leg is $(S^1\times I)\times S$. We can change the order of compactification and
regard the three space as $(S^1\times S)\times I$: in first step, we get a 4d ${\cal N}=4$ $SU(N)$ SYM
and we assume that the boundary condition at the ends of $I$ is classified by the same Young tableaux.
This fact can be seen from the following argument: the Hitchin's equation around the singularity
is identified with the Nahm's equation with specified singular boundary condition, which is
exactly the equation governing the boundary condition for ${\cal N}=4$ $SU(N)$ SYM on the half space.
In this order of compactification, 3d theories are represented as 4d ${\cal N}=4$ SYM on a one dimensional
graph. In fact, the graph is just the dual graph of the punctured Riemann surface
as described in \cite{dan1}, it is a trivalent graph with lots of three junctions.

With this graph representation of the theory $A$, the mirror symmetry is understood as the S duality of
the ${\cal N}=4$ on the graph. The S duality of ${\cal N}=4$ SYM on half space has been studied in full detail in \cite{GW1,GW2}, in particular,
the dual of the boundary condition we discussed earlier is worked out. The mirror of each boundary condition is a
quiver leg. For example, if the Young tableaux has heights $[h_1,h_2,...h_r]$ with $h_1\geq h_2\geq...\geq h_r$,
then the mirror quiver leg is
\begin{equation}
N-U(n_1)-U(n_2)-....-U(n_{r-1})
\end{equation}
where $n_i=\sum_{i+1}^rh_j$, and the first $N$ means we have a global $SU(N)$ flavor symmetry.

The S-dual of the three junctions is worked out in \cite{BYX}, it is simply the diagonal part of three $SU(N)$
gauge groups on the legs connecting with the junction. By combining various components, the mirror theory
is just a star-shaped quiver with a $SU(N)$ node at the center. In another word, we simply gauge
together the $SU(N)$ node of each leg. It is interesting to note that the mirror does not depend
on the pants decomposition.

\begin{figure}
\begin{center}
\includegraphics[width=4in]{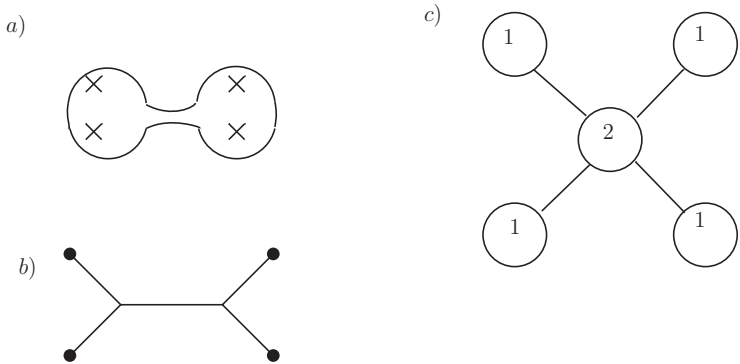}
\end{center}
\caption{(a)One S-duality frame for four dimensional ${\cal N}=2$ SU(2) with four fundamentals, each puncture carries
a SU(2) flavor symmetry. (b) Three dimensional version of (a), which is derived by compactifying (a) on a circle.
we represent it as $N=4$ SYM on the graph. (c)The graph mirror of (b), which is simply derived
by gluing the SU(2) flavor symmetry of four quiver tails.}
\label{mr1}
\end{figure}

Let's give an example to illustrate the main idea. Consider four dimensional ${\cal N}=2$
$SU(2)$ gauge theory with four fundamentals. It is derived from six dimensional theory
on a Riemann sphere with four punctures. One of the pants decomposition is described in fig.~\ref{mr1}(a).
The graph representation of 3d theory is shown in fig.~\ref{mr1}(b). The internal
leg represents the $SU(2)$ gauge group. The Young tableaux of the boundary condition has the
heights $[1,1]$, the mirror of this boundary condition
is a quiver tail $2-U(1)$, after gauging the common $SU(2)$ node, we found the mirror in fig.~\ref{mr1}(c).

We can extend the above analysis to $D_N$ theory \cite{SO, BYX}. Since the four dimensional gauge
theory involves not only $SO$ group but also USp group, we need to turn on the $z_2$ monodromy
line. In the pants decomposition, if there is a monodromy around the circle $S^1$, then the gauge group
on that leg would be $SO(2N+1)$. To get a three dimensional theory, we further compactify
the theory on $S$. To represent the theory as $N=4$ SYM on the graph, we change the order of
compactification, we first compactify
six dimensional theory on $S\times S^1$, the four dimensional gauge group is $USp(2N-2)$ which is
the $S$ dual of the theory derived from $S^1\times S$, the boundary condition at the ends should
be given by the Young tableaux of $USp(2N-2)$ which is exactly the case as found in \cite{SO}. There
are two types of junctions: the first type is the one for which there is no $USp$ leg while the
second type has two $USp$ legs. The S-dual of the boundary condition of $SO$ group and $USp$ group has
been also studied in \cite{GW2}. The S-dual of two different junctions are worked out in \cite{BYX}: the dual
is the diagonal $SO(2N)$ group for the first type of junction while the S-dual of the second type
is a diagonal $USp(2N-2)$ group. The dual of the boundary conditions are also worked out explicitly in \cite{BYX}.

\section{Adding more fundamentals}
\subsection{$A_{N-1}$ Theory}
   \subsubsection{Genus $0$ theory}
The theories studied in \cite{BYX} is superconformal in the four dimensional sense. It is interesting
to extend to the non-conformal cases, i.e. those theories with more fundamentals on the weakly coupled
gauge group(we call them theory $\tilde{A}$). We will use
the graph representation of the three dimensional theory we reviewed in last section. Before
doing that, we want to introduce some important concepts on 3d quiver gauge theories.

Since the IR theory we want to study is usually strongly coupled, we hope we can learn some of
its property from the UV theory, this is not always possible, for instance, there might
be accidental R symmetry in the IR which is not the same $R$ symmetry in the UV.
Consider  3d ${\cal N}=4$ $SU(N_c)$ theory with $N_f$ fundamentals, let's
define the excess number of it:
\begin{equation}
e=N_f-2N_c.
\end{equation}
This theory is called ``good'' if $e\geq 0$, ``ugly'' if $e=-1$, ``bad'' if $e<-1$. For the ``good'' theory,
there is a standard critical points and the IR R symmetry is just the R symmetry in the UV theory. For the
``ugly'' theory, the IR limit is just a set of free hypermultiplets. For the ``bad'' theory, the IR
limit is not a standard critical point and the R symmetry might be accidental symmetry. For the "good" theory, the
theory can be completely higgsed and there is a pure Higgs branch, we can learn a lot about the IR limit
from the UV theory. We mainly focused on the "good" theory in this section. If $e=0$, we
call it a ``balanced'' theory which has interesting property on the Coulomb branch.

The above definition can be extended to a quiver. We call a quiver ``good'' if $e_i\geq0$ for every node.
The Coulomb branch symmetry is enhanced due to the monopole operators. If we have a linear chain of
balanced quiver with $P$ nodes, i.e. $e_i=0$ for every node, then the global symmetry on Coulomb branch
is enhanced to $SU(P+1)$. If the balanced quiver has the shape $D_n$ or $E_n$ type dynkin diagram,
then the symmetry is enhanced to the corresponding $D_n$ or $E_n$ group. The global symmetry for a general ``good'' quiver
 is just the product of  enhanced non-abelian symmetries and abelian $U(1)$s from non-balanced nodes.
This is useful since we can read the exact global symmetry of Higgs branch of the theory $A$ using the mirror.
For instance, for the theory $SU(2)$ with four fundamentals, the flavor symmetry is $SO(8)$.
In the Gaiotto's representation in fig.~\ref{mr1}(a), only $SU(2)^4$
subgroup is manifest, while we can see the full $SO(8)$ symmetry in the Coulomb branch of the mirror using monopole
operators. This example might be trivial since we have a lagrangian description for $A$,
 but for other strongly coupled theory, the mirror theory is very useful to see full flavor symmetries.

For the irreducible theory $A$ we considered in this paper, the mirror $B$ is always good as one
can check. Now let's consider theory $\tilde{A}$ which is derived by adding more fundamentals to the gauge groups of the theory
$A$ considered in \cite{BYX}, the mirror $\tilde{B}$ should also be a good quiver. There is a graph representation
for $A$ as we reviewed in last section, the gauge groups are represented as the internal legs. To add fundamentals,
it is natural to think adding some ``D5'' branes on the internal leg. This is really a Type IIB language, let's
go back to Type IIA or M theory language and the new system can be similarly realized as the six dimensional construction
\footnote[1]{We thank referee of raising this question.}. In the weakly coupled limit, the long tube region is locally as
a manifold $S^1\times I \times S \times R^3$ as reviewed in last section,  adding more fundamental in type IIA language
means adding more $D6$ branes in Witten's brane construction, in lifting to M theory, $D6$ branes becomes Taub-Nut space,
so the three dimensional theory is realized as  $M5$ branes compactified on a punctured Riemann surface in Taub-Nut
space and then further compactified on a circle. However, we find the type IIB language is more useful
in finding 3d mirror since one can use the knowledge of field theory.

There is one important question:
Is the IR limit of ${\cal N}=4$ SYM on the graph the same as the IR limit of theory $\tilde{A}$?
In general the answer is not: there are some free hypermultiplets in the IR besides the fix point theory of
$\tilde{A}$. This can be seen from the mirror of the graph. The graph mirror is in general a ``bad'' quiver.

\begin{figure}
\begin{center}
\includegraphics[width=4in]{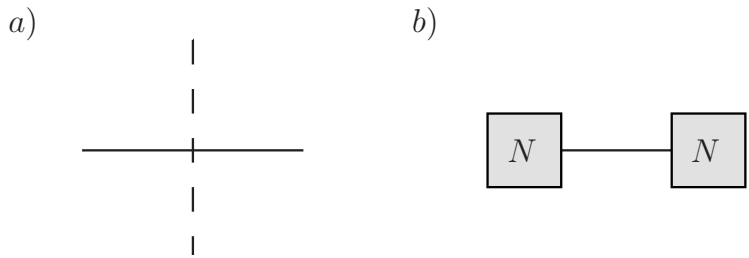}
\end{center}
\caption{(a) The addition of a ``D5'' brane to the internal leg of the graph. (b)Its mirror.
}
\label{mr2}
\end{figure}

The mirror of the graph is simple, the S-dual of the ``D5'' branes are ``NS'' brane. From the gauge theory point
of view, there are now two $U(N)$ gauge groups connected by a bi-fundamental, see fig.~\ref{mr2}. In general, the graph mirror
is ``bad'' which reflects the fact the IR limit of the graph does not coincides with the IR limit of theory $\tilde{A}$.
We want to extract the mirror $\tilde{B}$ from the graph mirror.

The process we suggest is the following: for any ``bad'' or ``ugly'' node on the graph mirror with the excess number $e_i<0$,
we replace its rank by
\begin{equation}
n_i^{'}=n_i+e_i=N_f-N_c,
\end{equation}
then the excess number of the quiver nodes around it will also be changed,
if there are still some ``bad'' nodes, we will do the same manipulation on those nodes. We continue doing this until
all the nodes are ``good'', the resulting theory is the mirror $\tilde{B}$ for the theory $\tilde{A}$ (The overall
$U(1)$ of $\tilde{B}$ is decoupled and we do not include this factor in counting the Higgs and Coulomb branch).

\begin{figure}
\begin{center}
\includegraphics[width=2in]{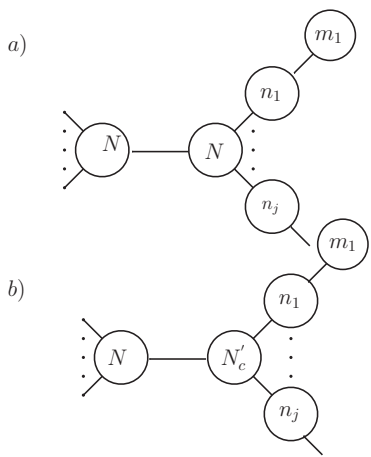}
\end{center}
\caption{(a) The naive mirror for adding one fundamental to one of weakly coupled gauge group; We
assume that the left central node is bad.
(b)We replace the rank of the left central node with $N_c^{'}=N_f-N=\sum_{k=1}^jn_k$.
}
\label{mr3}
\end{figure}

The theory $\tilde{B}$ should have the same Higgs branch as the star-shaped quiver $B$. Since theory $\tilde{A}$ has
the same Coulomb branch dimension as $A$. The graph mirror
has the same Higgs branch as $B$ as we can easily count: we add one bi-fundamental and one $U(N)$ node, the net
contribution to Higgs branch is zero. The graph mirror has two central nodes, and the only possible
``bad'' node is one of the central node, See fig.~\ref{mr3}. for the illustration. The contribution of this central node
to the Higgs branch is
\begin{equation}
N_fN_c-N_c^2=N_c(N_f-N_c).
\end{equation}
This number is unchanged if we change the rank of the gauge group to
$N_c^{'}=N_f-N_c$ and keep $N_f$  unchanged. The number $N_c^{'}=N_c+N_f-2N_c=N_c+e$ which is just the number
we defined earlier. $N_c^{'}$ should be less or equal than $N$, so we can only do
the manipulation for those quiver node with $e<0$.

We want to point out some generic features of the manipulation. As we noticed earlier, only one of the
central node can be ``bad'' for the graph mirror, the excess number of it is $e<0$. After changing its rank, its new
excess number is $-e$. The excess number of its adjacent nodes are increased by $e$. If none of
those new excess numbers are negative, then our manipulation stops, there is one more $U(1)$ global
symmetry on the Coulomb branch from the new node. This reflects the fact there is an extra $U(1)$
flavor symmetry coming from the new added fundamental. It is possible some of the adjacent node
becomes ``balanced'' and therefore we have enhanced symmetry, however we can only have one new ``balanced'' adjacent
node with just one exception. We order the rank of the adjacent nodes so that $n_1\leq n_2\leq...\leq n_j$,
so
\begin{equation}
e_i^{'}\geq \sum_{k=1}^jn_k-2n_i,
\end{equation}
$e_i^{'}\geq0$ for any $i\geq3$. $e_2\geq0$, it is zero only when there are only two identical quiver tails with
Young tableaux $[N-n_1,n_1]$. This is consistent since after adding only one fundamental, the flavor symmetry
can only be changed from $SU(k)$ to $SU(k+1)$ or from $SO(k+2)$ to $SO(2k+2)$ (this is for the $USp$ group).

It is possible that $n_1$ node becomes  ``bad'' after we change the rank of the central node.
We will focus on this quiver tail. The excess number of each node on this quiver tail can be read
from the Young tableaux:
\begin{equation}
e_i=n_{i+1}+n_{i-1}-2n_i=\sum_{j=i+2}^{r}h_j+\sum_{j=i}^{r}h_j-2\sum_{j=i+1}^rh_j=h_{i}-h_{i+1},
\end{equation}
we take $h_{r+1}=0$ (here we use $i$ to denote the node on this particular quiver tail and $n_i$ as its rank), the excess number
is non-negative as from the definition of the Young tableaux. It is a several chains of balanced quiver
separated by the ``good'' quiver nodes.

After changing the rank of this node, the new excess number of the central number is $e_c^{'}=e_1$ which
is positive. The excess number of the $n_1$ is $e_1^{'}=-e_1-e$.
The new excess number of other adjacent node $n_2$ is $e_2^{'}=e_1+e_2+e$, if this number is non-negative,
then our process stops. If not, we continue the process, the excess number of the first node changed to $e_2$ though.
We only need to do one manipulation on each possible node. The general conclusion is that the process stops at
the $j$th node on the quiver tail with the condition
\begin{equation}
e_1+e_2+...e_j+e\geq 0,~~e_1+e_2+..e_{j-1}+e<0. \label{rank}
\end{equation}
In particular, the structure of the balanced chain is not changed. The final form of the quiver with its rank and
excess number is shown in fig.~\ref{mr16}. No new rank number is zero or negative, since
\begin{equation}
n_{i}^{'}=n_i+e+e_1+...e_i=n_i+h_1-h_{i+1}+e>n_i+h_1-h_{i+1}+n_1-N=n_i-h_{i+1}>0.
\end{equation}
This ensures that no quiver node disappears. The structure of the new quiver shows that there is a extra $U(1)$ on
the symmetry of the Coulomb branch, which is exactly what we want.

\begin{figure}
\begin{center}
\includegraphics[width=4in]{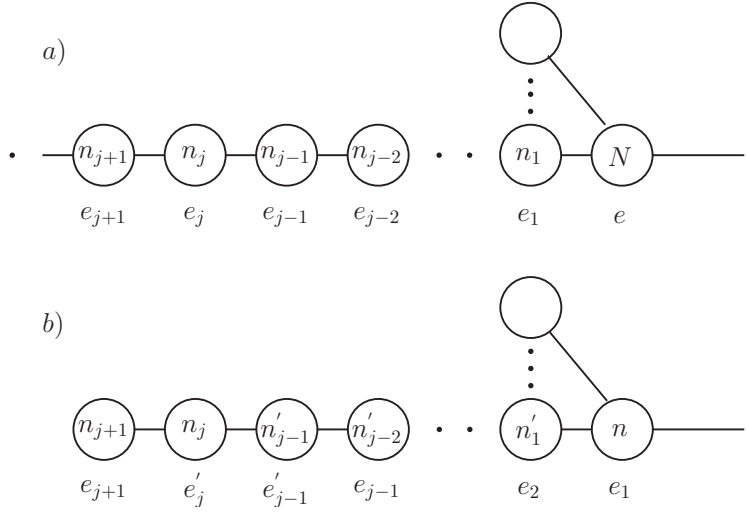}
\end{center}
\caption{(a) The rank and excess number of a quiver tail associated with a central node, we assume that $e+e_1<0$ for this
quiver tail. (b)The rank and excess number of the quiver tail after the manipulation is finished. It stops at the $j$th node, from
the condition, we have $e_j>0$, this shows that no balanced node is lost, and they balanced chain is not altered.
The new excess number is $e_{j-1}^{'}=-(e+e_1+e_2+...e_{j-1})>0$, $e_{j}^{'}=e+e_1+e_2+...e_j\geq 0$, so it is only possible for one more
balanced node to appear. If a new balanced node appears, it shows that there are already fundamentals exist;
The new rank is $n_{i}^{'}=n_i+e+e_1+...e_i$.
}
\label{mr16}
\end{figure}

For four dimensional theory, the gauge group contents depend on the pants decomposition of the Riemann surface;
To add fundamentals to the gauge group, we must specific the pants decomposition. Go to three
dimensions, the mirror is obviously different for different pant decomposition,
which is in contrast with conformal case for which the mirror is independent of pants decomposition. This
allows us to determine different duality frames of 4d SCFT as we will see later.

\begin{figure}
\begin{center}
\includegraphics[width=2.5in]{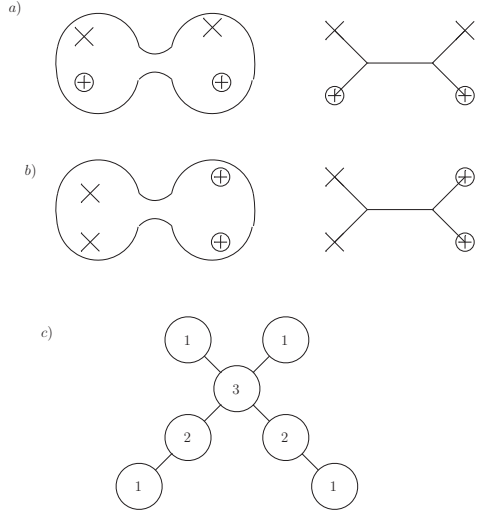}
\end{center}
\caption{(a)The weakly coupled duality frame with SU(3) gauge group on the left, there are two type of punctures: the cross
represents the simple puncture with Young tableaux $[2,1]$, the circle cross represents the full puncture with tableaux
$[1,1,1]$. The graph representation for
three dimensional theory is shown on the right. (b) The weakly coupled duality frame with SU(2) gauge group on the left,
graph representation for three dimensional theory on the right. (c) The mirror for theory (a) and (b), they are identical.}
\label{mr4}
\end{figure}

Let's give an example to illustrate our main idea. The four dimensional theory is the original example studied
by Argyres-Seiberg \cite{Argy}, in one weakly coupled duality frame $A_1$,  it is just a $SU(3)$ theory with six fundamentals;
In another duality frame $A_2$, there is a weakly coupled $SU(2)$ gauge group coupled with one fundamental and $E_6$
strongly coupled theory. The S-duality can be understood from the six dimensional construction. See fig.~\ref{mr4} for
the pants decomposition and the graph representation for the corresponding three dimensional theory.

\begin{figure}
\begin{center}
\includegraphics[width=4in]{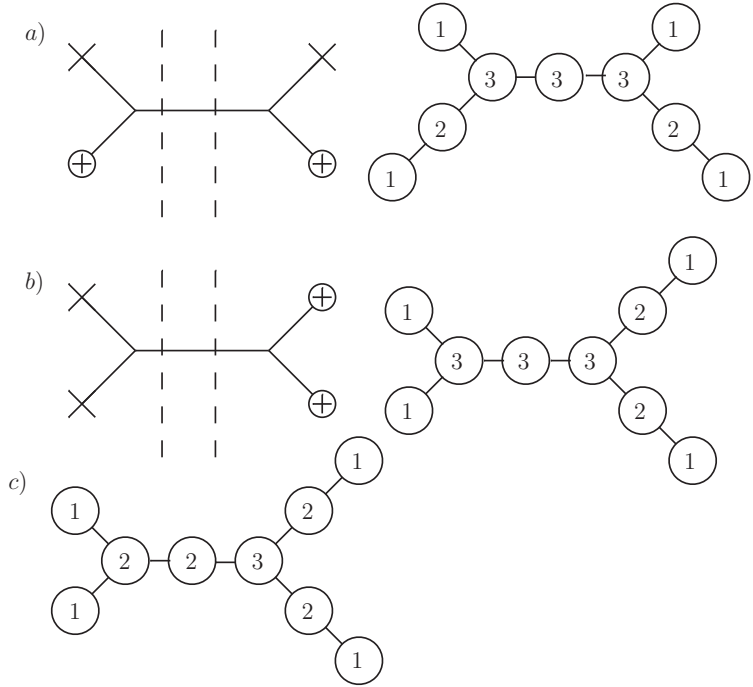}
\end{center}
\caption{(a)On the left, we add two ``D5'' branes on the internal segment which represents a SU(3) gauge group,
 on the right, we apply S-duality and get a good quiver.
(b)We add two ``D5'' branes on the $SU(2)$ gauge group, the left-hand side is the mirror quiver which is bad,
the left central $U(3)$ node is ``bad''.
(c)We replace the bad $U(3)$ node with $U(2)$, and then replace the adjacent $U(3)$ with $U(2)$ node, the
resulting quiver is ``good''.
}
\label{mr5}
\end{figure}

Now let's add more fundamentals (say two as in fig.~\ref{mr5}) to the gauge groups of the above theories, namely, we now consider
theories $\tilde{A}_1$ and $\tilde{A}_2$. To find their mirrors, we use the graph representation of the
conformal theories and add ``D5'' branes on the internal leg, we apply the S-dual and find graph mirror. If
the mirror quiver is ``good'', then this quiver is just the mirror of $\tilde{A}$; if the mirror is ``bad'', this means
that the IR limit of the graph is not the same as the theory $\tilde{A}$, but we can do the manipulation as
we described earlier to find the mirror of $\tilde{A}$.

The graph representation and graph mirrors  are shown in fig.~\ref{mr5}. The simple puncture is represented
by the Young tableaux with heights $[2,1]$, the quiver tail to it is just $3-U(1)$; The circle cross has
partition $[1,1,1]$, the quiver tail is $3-U(2)-U(1)$. The mirror of ``D5'' branes is to cut the gauge groups into two and
introduce a bi-fundamental connecting them.

The quiver in fig.~\ref{mr5}(a) is ``good'', so we conclude that the graph representation
on the left of fig.~\ref{mr5}(a) has the same IR limit as the three dimensional
$SU(3)$ with 8 fundamentals, and the graph mirror is just the mirror of theory $\tilde{A}_1$,
this is in agreement with the result in \cite{HW}.

The quiver in fig.~\ref{mr5}(b) is ``bad'': the SU(3) node on the left has excess number negative 1, so we replace it
with a $U(2)$ node, then the SU(3) node adjacent to it becomes ``ugly'' with excess number negative 1, we also replace
it with $U(2)$. After doing this, we get a "good" quiver as shown in fig.~\ref{mr5}(c), this is the mirror for the theory $\tilde{A}_2$.

Let's do some check on our result. The theory $\tilde{A}_2$ has the same Coulomb branch dimension (we always mean the hyperkahler
dimension in this paper) as $A_2$ and the Higgs
branch dimension of $\tilde{A}_2$ is increased by four. Comparing the quiver in fig.~\ref{mr5}(c) with the quiver in fig.~\ref{mr4}(c), its Coulomb
branch dimension is increased by 4 and Higgs branch dimension is not changed. The flavor symmetry of $\tilde{A}_2$
is $SO(6)\times SU(6)$. The $SO(6)$ is from three fundamentals while $SU(6)$ is from $E_6$ matter. In the fig.~\ref{mr5}(c), on the
left, we have a linear chain of three balanced quiver and on the right we have a linear chain with 5 balanced quiver, so the
symmetry on the Coulomb branch is $SU(4)\times SU(6)$ which is the same as the symmetry on the Higgs branch of $\tilde{A}_2$.
(Notice the $U(1)$ symmetry on the middle $U(2)$ node is decoupled).

\begin{figure}
\begin{center}
\includegraphics[width=4in]{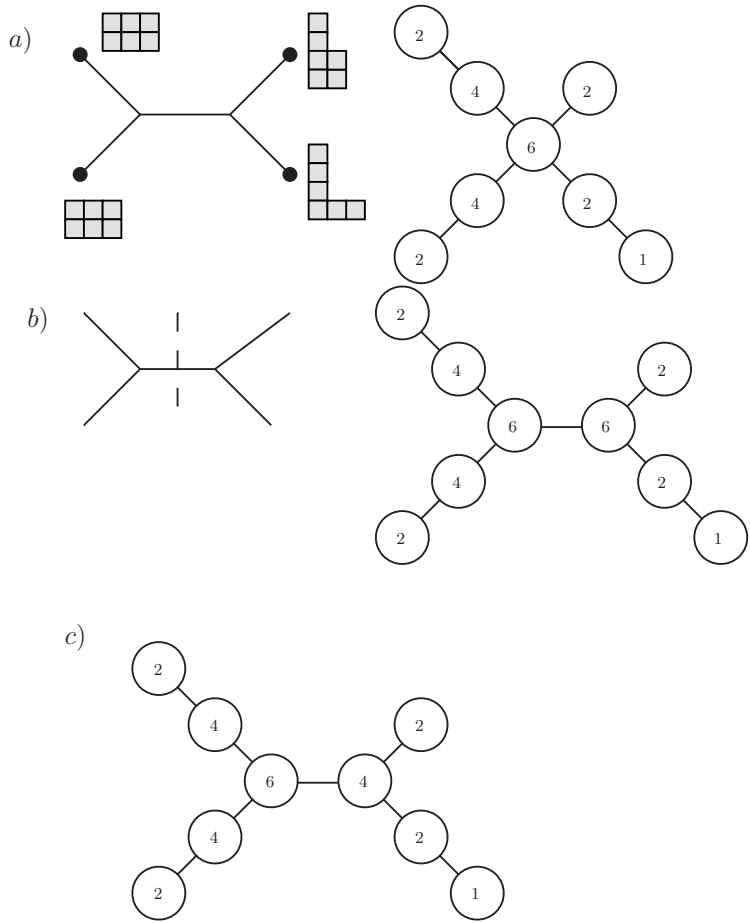}
\end{center}
\caption{(a)A generalized quiver gauge theory $A$ in three dimensions, its graph representation is depicted on
the left, we draw Young tableaux for the boundary condition, its mirror is depicted on the right.
(b)We add one ``D5'' branes on internal leg of the graph in (a), the mirror of the graph is shown on the right.
(c)The mirror of the theory $\tilde{A}$ which has one more fundamental than $A$.}
\label{mr6}
\end{figure}

In fact, we can use the ``D5'' brane as a probe to find out what is the weakly coupled gauge group $SU(k)$ (or $USp(k)$
in some cases) for 4d SCFT by counting the change of the Coulomb branch dimension of the mirror. Since for the theory $\tilde{A}$, the
Higgs branch is increased by $k$, if we know the change of Coulomb branch dimension of the mirror, we can
 determine the weakly coupled gauge group. To determine whether it is a $USp$ group of $SU$ group, we can
see the enhanced symmetry on the Coulomb branch of the mirror. If the mirror
quiver has a balanced part with shape of $D_n$ dynkin diagram, then the gauge group is $USp(k)$,
otherwise the weakly coupled gauge group is $SU(N)$.
The weakly coupled gauge group can also be determined  using the degeneration limit in \cite{dan1}.
Here we  use three dimensional mirror symmetry to do the job. We describe one example in fig.~\ref{mr6}
Comparing the quiver in fig.~\ref{mr6}(c) and fig.~\ref{mr6}(a), the Coulomb
branch dimension is increased by $4$, since the quiver does not have a balanced part with $D_n$ dynkin diagram shape,
 the gauge group is $SU(4)$, this is in agreement with the result using the degeneration method as described
in \cite{dan1}. A special case is if the graph mirror is a ``good'' quiver, then the gauge group is $SU(N)$
or $USp(N)$ as the Coulomb branch of the mirror is increased by $N$.

We can also find out how many fundamentals on the gauge group for the four dimensional conformal theory.
Since if there are $l$ fundamentals exists, after adding one more fundamental, the global symmetry on Higgs
branch is enhanced from $U(l)$ to $U(l+1)$. In the mirror, we can see the change of the global symmetry
on Coulomb branch using monopole operators, and we can determine $k$. In the quiver of fig.~\ref{mr6}(a), the global symmetry
on Coulomb branch is $SU(6)\times U(1)\times SU(2)$; For the quiver in fig.~\ref{mr6}(c),
the symmetry on Coulomb branch is $SU(6)\times SU(2) \times U(1) \times SU(2)$, we see the global
symmetry is changed from $U(1)$ to $U(2)$, so originally we have only 1 fundamental. This is also in
agreement with the result in \cite{dan1}.

\begin{figure}
\begin{center}
\includegraphics[width=4in]{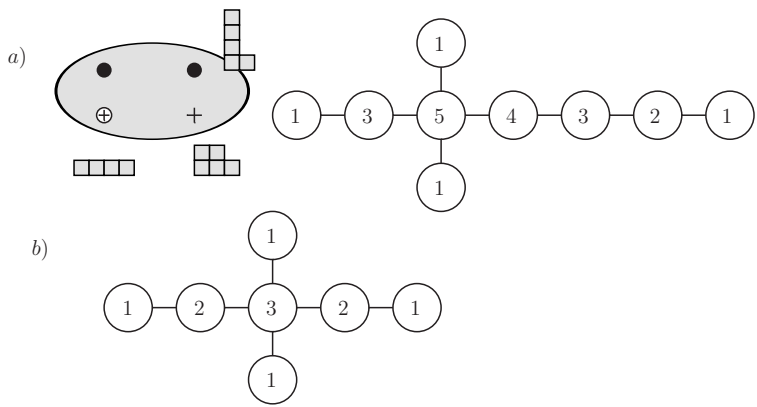}
\end{center}
\caption{(a) A four dimensional theory is derived by compactifying six dimensional $A_4$ theory on
a sphere with four punctures, the Young tableaux of the punctures are depicted, its 3d mirror
can be found from the information in the puncture, it is a bad quiver, the central node
has negative excess number. (b) For the bad node, we change its rank using the formula $N_c^{'}=N_c+e$,
after this modification, the excess number of the quiver nodes around the central are also changed and
they are ``bad'', we do the modification on those nodes too, at the end, we get a ``good'' quiver, this quiver
is the same as the quiver depicted in figure 4.}
\label{mr7}
\end{figure}

\begin{figure}
\begin{center}
\includegraphics[width=3.5in]{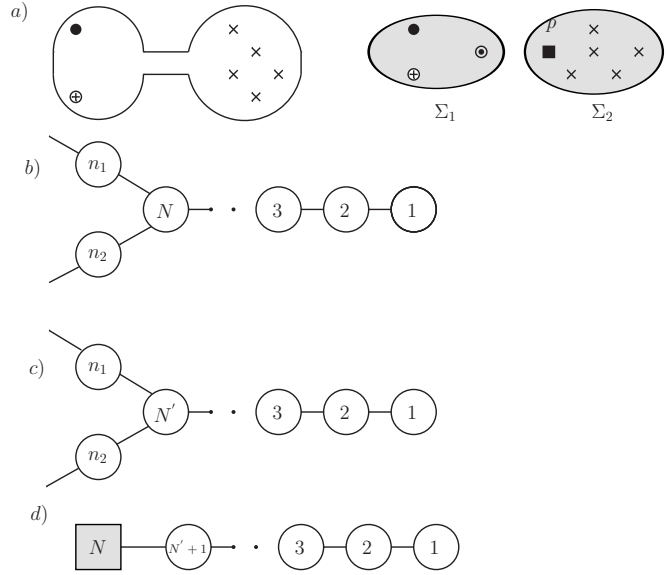}
\end{center}
\caption{(a) A four dimensional ${\cal N}=2$ generalized quiver gauge theory is derived from six dimension $(0,2)$ $SU(N)$
SCFT on this punctured sphere, the gauge group is represented by a long tube;
We show one weakly coupled gauge group at the end of the quiver; After completely decoupling the gauge group, the
Riemann surface becomes into two parts $\Sigma_1$ and $\Sigma_2$, we put a full puncture at the new formed three punctured sphere. There is
a new puncture $p$ on $\Sigma_2$.
(b) The 3d mirror for three punctured sphere. (c) If the mirror in (b) is ``bad'', we change the rank of the ``bad'' nodes and get
a ``good'' quiver, we assume that $n_1$ and $n_2$ nodes are ``good''. (d) The mirror quiver tail for the puncture $p$ for the
case $(c)$.}
\label{mr17}
\end{figure}

\begin{figure}
\begin{center}
\includegraphics[width=3in]{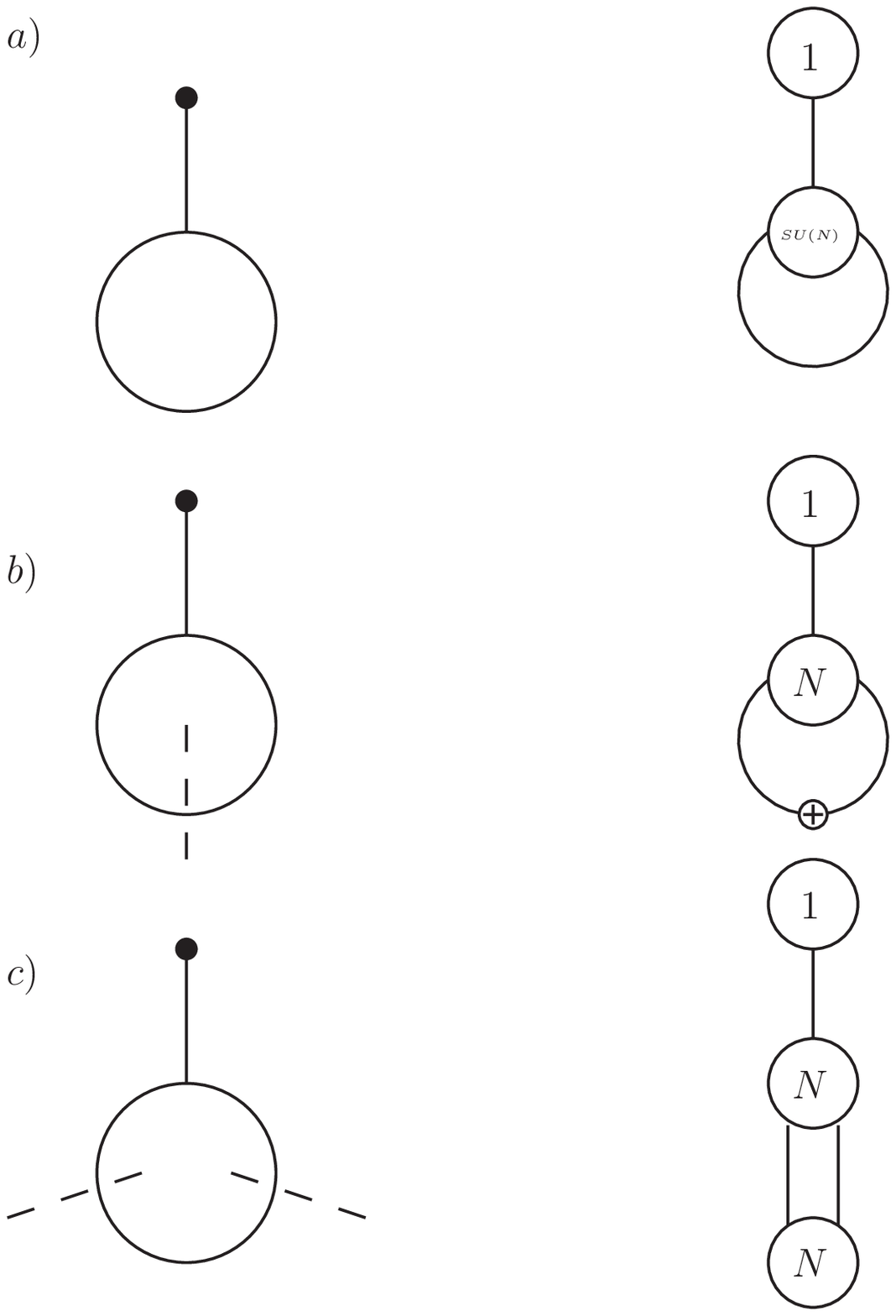}
\end{center}
\caption{(a)The graph representation of ${\cal N}=2$ $SU(N)$ with an adjoint is depicted on the left,
the mirror theory is depicted on the right, the loop attached on $SU(N)$ is the adjoint of SU(N).
(b)On the left, we add one more fundamental to theory a); There is a cross on the adjoint which
means that here is the adjoint on U(N), also the central node is $U(N)$ group.
(c)We add two fundamentals to theory a); The mirror is depicted on the right, no $U(1)$ is projected out.
}
\label{mr8}
\end{figure}

There is another application of our procedure of extracting irreducible theory from ``bad'' quiver.
We consider four dimensional irreducible theory up to now. For the reducible theory, the three
dimensional mirror is ``bad'', which means that there are free hypermultiplets besides the
SCFT, those SCFT are actually represented irreducibly by lower rank six dimensional $(0,2)$ theory.
We hope to extract the free hypermultiplets and the irreducible SCFT from the 3d mirror using the surgery
of the naive mirror. However, there is one point we want to clarify \footnote[2]{We thank referee point
out this issue and this motivates the following discussion.}. The rational for us to do the surgery 
is that the original theory has two parts $A+free$, where $A$ is a theory with non-zero Coulomb branch, 
the naive mirror $B$ is the mirror for
$A+free$. Obviously, the Higgs branch of $B$ should be the same as the Coulomb branch of $A$, our 
surgery ensures that the Higgs branch of the new quiver after the surgery has the same dimension as 
$B$. Finally, by comparing the Coulomb branch of $B$ and $B^{'}$, we can find the number of free 
hypermultiplets in the original theory. 

The above surgery works well if the original theory has non-trivial Coulomb branch which is 
the case we considered so far. However, there are some theories $A$ for which there is no 
Coulomb branch for example a sphere with three simple punctures, in this case, the number of 
free hypermultiplets are just the Coulomb branch dimension of the naive mirror. To check 
whether a theory has only free hypermultiplets, we do the surgery on the naive mirror, and
at some stage, the rank of gauge group after the surgery will be negative.

Those reducible theories have been considered in \cite{dist}, our method gives another
simple way to study them. For instance, consider $A_4$ theory on a sphere with four punctures which are labeled as
$[4,1],[4,1], [1,1,1,1],[2,2,1]$ (This is the first example in the appendix of \cite{dist}).
The 3d mirror of this theory is depicted in fig.~\ref{mr7}(a), it is a bad quiver, so we do the
modification to the rank of the bad quiver node, finally, we get a good quiver which is
depicted in fig.~\ref{mr7}(b). One can recognize that this good quiver is the same as the quiver in
fig.~\ref{mr4}, so we can conclude that this is the SCFT part of the theory. Comparing the quiver
in fig.~\ref{mr7}(a) and fig.~\ref{mr7}(b), the Coulomb branch is decreased by 10. So we conclude that the theory composes
of a SCFT as described in fig.~\ref{mr7} and 10 free hypermultiplets. This is in agreement with
the result in \cite{dist}.

Let's go back to four dimensional ${\cal N}=2$ generalized quiver gauge theory. We have shown how
to determine the weakly coupled gauge group and the number of fundamentals on it by using
the ``D5'' brane probe. However, there are other strongly coupled matter systems coupled with the gauge group,
we want to determine them. Let's follow the procedure in \cite{dan1}: the weakly coupled gauge group corresponds
to the long tube of the Riemann surface; we first consider the gauge group at the end of the quiver and
completely decouple this gauge group, two new punctures appear,
 the Riemann surface are decomposed into two parts:  a three punctured
sphere $\Sigma_1$ and another sphere $\Sigma_2$ with a lot of punctures, see fig.~\ref{mr17}(a). The information of the matter system
can be read from the three punctured sphere, we also want to determine what is the new puncture $p$ on $\Sigma_2$.
In \cite{dan1}, we assume that the two new appearing punctures are identical and find the new puncture by counting
the Coulomb branch dimension. Motivated by our study of mirror symmetry in this paper, we follow a different approach,
we assume the new puncture on $\Sigma_1$ is always the maximal puncture.

We extract the matter information by looking at the 3d mirror of the three punctured sphere $B$, see fig.~\ref{mr17}(b) , If the mirror is ``good'', then the matter system is just a strongly coupled isolated SCFT and the puncture $p$ is just the maximal puncture. The
weakly coupled gauge group is just $SU(N)$ as we have shown earlier. If the mirror quiver is ``bad'', we
can still extract the matter information by doing the manipulation we have used to find the mirror for the theory
with more fundamentals. There are several situations we need to consider. The first situation is that after changing
the rank of the central node to $n_1+n_2-1$, the $n_1$ and $n_2$ nodes are ``balanced'' or ``good'', in this case, the
resulting quiver $M_3$ is shown in fig.~\ref{mr17}(c). This is the matter coupled to the gauge group. The quiver tail
for the new puncture $p$ is shown in fig.~\ref{mr17}(d). There are some checks on our result. The weakly coupled
gauge group is $SU(n_1+n_2)$ from our previous analysis; For the quiver fig.~\ref{mr17}(d), there is indeed
a chain of $n_1+n_2-1$ ``balanced'' nodes which has an enhanced $SU(n_1+n_2)$ symmetry on Coulomb branch.
The original quiver is formed by gauging this global symmetry and the $SU(n_1+n_2)$ symmetry of the quiver $p$.
Another serious check is to compare the Coulomb branch and Higgs branch dimension of the decomposed system and
generalized quiver, they are in agreement with each other (the calculation is the same as we have done in \cite{dan1},
though a little bit tedious).

The other cases are more complicated. One usually have both free fundamental hypermultiplets and strongly
coupled matter system. We already know how to see fundamentals, using the above method one can also
extract the strongly coupled matter system.

\subsubsection{Higher genus theory}
Let's next consider the theory associated with the higher genus Riemann surface. The theory $A$ is a generalized
quiver gauge theory and the mirror theory $B$ has adjoint matter attached to the central $SU(N)$ node. There
is no complete Higgs branch for the $A$ theory: for genus $g$ theory, there are $g(N-1)$ free $U(1)$
vector hypermultiplets in the Higgs branch. So the dimension of the ``Higgs'' branch is equal to the number
of hypermultiplets minus the vector multiplets, and then plus $g(N-1)$. Let's
consider genus one Riemann surface with just one simple puncture, the graph and the mirror is shown in fig.~\ref{mr8}(a).
This is just 4d ${\cal N}=2^*$ $SU(N)$ theory, one should be a little bit careful here, the adjoint matter has dimension
$N^2$. One can check that the Coulomb branch and Higgs branch dimensions of $A$ and $B$ matches by counting the dimension of
``Higgs'' branch of $A$ by including the free $U(1)$ vector multiplets.

There are three types of internal legs for the higher genus theory in the dual graph. For the first one,
when we cut it, the number of loops of the graph is not changed.
When we cut the second type of internal legs, the graph becomes two disconnected parts with loops; when we
cut the third type of internal legs, there is only one part left and its number of loops is reduced by 1.
When we add more fundamentals to the first two types of internal legs, the procedure of finding the mirror
is the same as the genus zero case. In particular, for the second type of internal legs, the gauge group
must be $SU(N)$, adding a fundamental just introduces another U(N) group and a bi-fundamental.

For the third type of internal legs, there is a subtle point when we add just one fundamentals.
Consider the example in fig.~\ref{mr8}, we add a ``D5'' brane on
the internal leg as depicted in fig.~\ref{mr8}(b). Do S-dual on the graph, the ``D5'' brane becomes a ``NS5'' brane, which cuts the original
$U(N)$ group into two $U(N)$ groups, however, these two $U(N)$ groups are connected by a single junction which
must be a single $U(N)$ since in the mirror only the diagonal part of the junction is survived. The mirror
has just one $U(N)$ node. We need one modification: we replace the adjoint of $SU(N)$ with adjoint of
$U(N)$, and we let the central node be $U(N)$ instead of $SU(N)$. See fig.~\ref{mr8}(b) for the mirrors.

Let's compare the Coulomb branch and Higgs branch dimension of theory
$\tilde{A}$ and $\tilde{B}$. Now the Higgs branch of $\tilde{A}$ is just the difference between the
hypermultiplets and vector multiplets. It is easy to see the dimensions matches. There are
two mass parameters for the theory $\tilde{A}$ and in the mirror there are two $U(1)$ factors so we
have two FI parameters as it should be.

When we add two fundamentals, there is no adjoint in the mirror, we have two $U(N)$ gauge groups,
see fig.~\ref{mr8}(c). In general, when we add $k$ fundamentals, there are $k$ $U(N)$ gauge groups in the mirror.

In general, for genus $g$ theory, when we add just one fundamental on one of the handle, in the mirror,
the central node is changed to $U(N)$ and one of the adjoint of $SU(N)$ is changed to the adjoint of $U(N)$; After doing
that, there is only $(g-1)(N-1)$ free $U(1)$s in the ``Higgs'' phase of $\tilde{A}$. If we
add another fundamental to a different handle, we simply change one of the adjoint of $SU(N)$ to $U(N)$,
the number of free $U(1)$ is reduced by $(N-1)$.
However, when we add one fundamental on each handle, there is not enough FI terms in the mirror,
what happens we believe is that there is hidden ``FI'' terms which appear only in the IR. There are a total
of $(g-1)$ hidden ``FI'' terms.

\begin{figure}
\begin{center}
\includegraphics[width=4in]{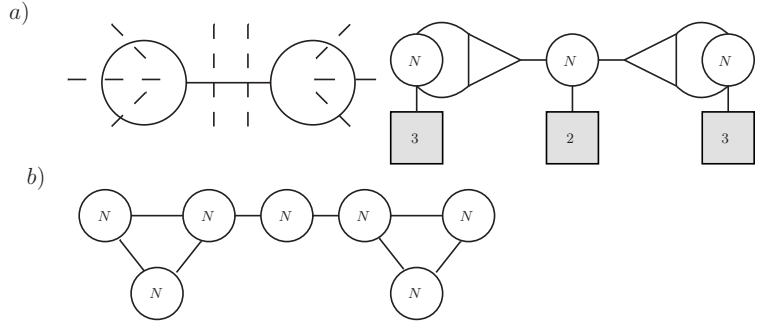}
\end{center}
\caption{(a)A genus two theory with several fundamentals, the generalized quiver representation is shown on the left
. (b) Mirror theory of (a).
}
\label{mr9}
\end{figure}

We can add arbitrary number of fundamentals to any of the weakly coupled gauge group.
A genus two example is shown in fig.~\ref{mr9}.

\subsection{$D_N$ theory}
The above analysis can be extended to $D_N$ theory. Let's first discuss the definition of the ``good'',
``bad'', ``ugly'' for the $USp$ and $SO$ gauge theory. For $SO(k)$ gauge theory with $n_f$ flavors, we define
the excess number
\begin{equation}
e=n_f-k+1.
\end{equation}
The theory with $e\geq 0$ is called ``good''. For $USp(2t)$ theory with $n_f$ flavors, the excess number is defined as
\begin{equation}
e=n_f-2t-1.
\end{equation}
The theory with $e\geq 0$ is ``good''. The above theories are called ``balanced'' if $e=0$, notice that the balance
condition for three dimensional theory is different from the conformal invariant condition for the
four dimensional theory. So the conventional conformal orthosymplectic quiver is not a ``good'' quiver,
which is different from the unitary case. In fact, the $SO$ nodes are ``bad''.

Similarly, a quiver with alternative $USp$ and $SO$ nodes
are called ``good'' quiver if $e_i\geq 0$ for every node
in the quiver; It is called ``balanced'' if $e_i=0$ for every node. The global symmetry of Coulomb branch is enhanced by monopole operators
for a chain of balanced orthosymplectic quiver with P nodes, the global symmetry is in general enhanced to $SO(P+1)$. However,
if the first node on the chain is $SO(2)$, the global symmetry on the Coulomb branch of a chain with $P$ nodes  is $SO(P+2)$.

For the theory considered in \cite{BYX}, there is $USp$ global symmetry for the $A$ theory, but there is no balanced
orthosymplectic quiver with enhanced $USp$ flavor symmetry, so in general the mirror quiver is a ``bad'' quiver.

\begin{figure}
\begin{center}
\includegraphics[width=4in]{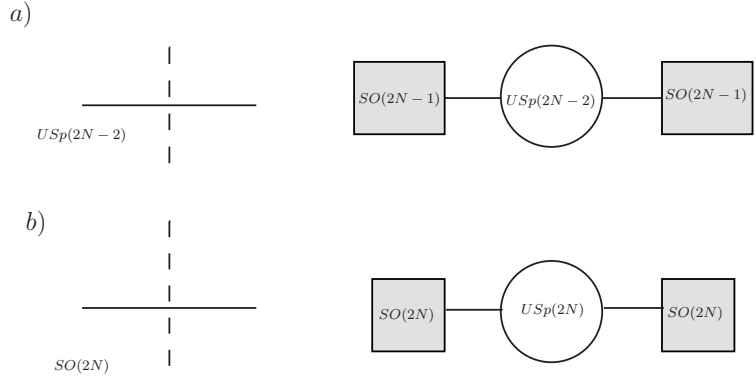}
\end{center}
\caption{(a) The addition of a ``D5'' brane to a USp leg, its mirror is depicted on the right;
(b) The addition of a ``D5'' brane to a SO leg and its mirror.
}
\label{mr10}
\end{figure}

There are two types of internal legs for the theories considered in \cite{BYX}. We add
some full ``D5'' branes to the internal leg and the mirror of one ``D5'' brane is shown in \cite{GW2} using
brane splitting, we reproduce it in fig.~\ref{mr10}.

\begin{figure}
\begin{center}
\includegraphics[width=6in]{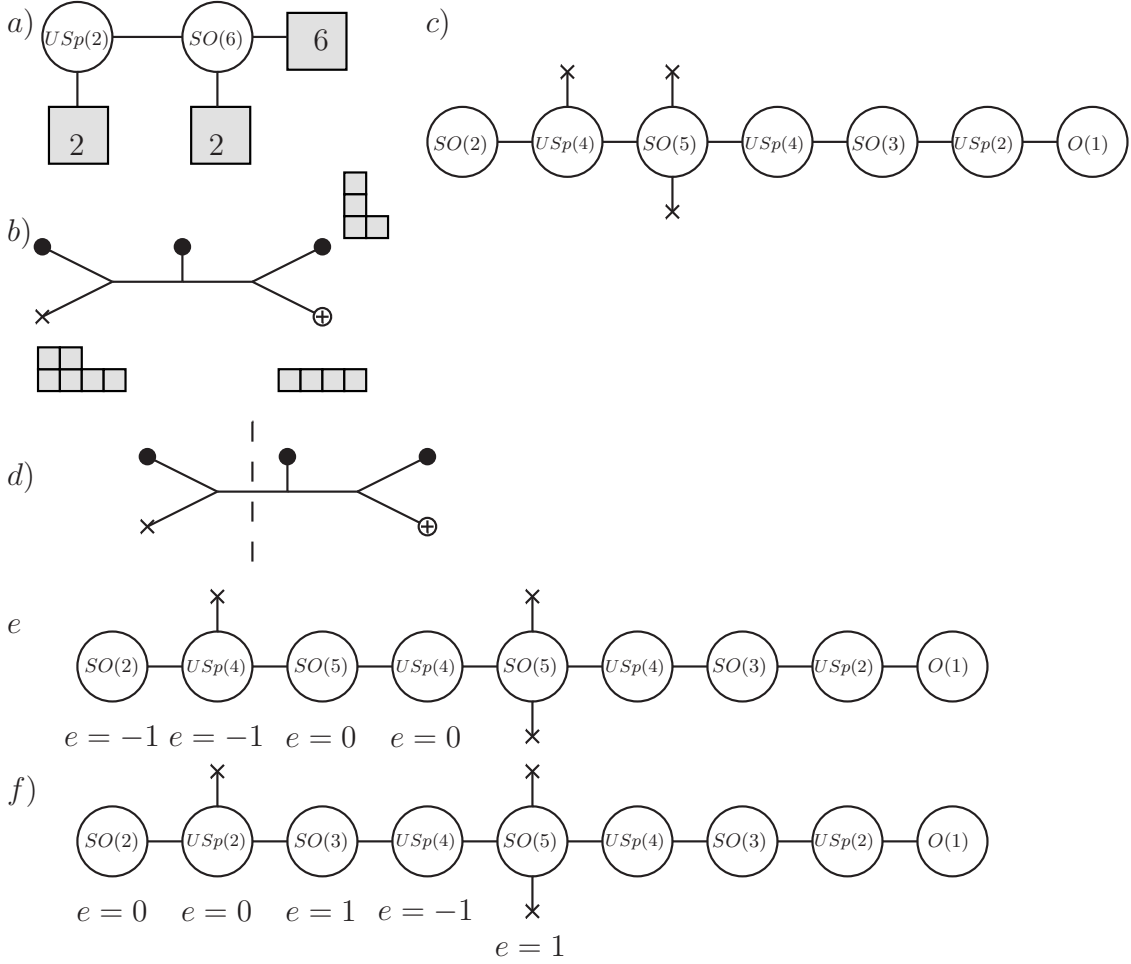}
\end{center}
\caption{(a)A orthosymplectic quiver which is conformal in four dimension. (b)The graph representation of
the three dimensional theory; We indicate the Young Tableaux here. (c) The mirror theory $B$.
(d) We add one full ``D5'' brane on the internal leg on $USp(2)$ leg. (e) The naive mirror of the (d), we
indicate the excess number for some of the relevant nodes; We can do the rank manipulation on $USp(4)$ node.
(f) we change the $USp(4)$ node to $USp(2)$,
and then change the adjacent $SO(5)$ node to $SO(3)$, the new excess number is indicated. This is
the mirror theory $\tilde{B}$.
}
\label{mr11}
\end{figure}

To find the mirror theory $\tilde{B}$, we can not simply extend our analysis for the unitary group case,
since the mirror $B$ is already a bad quiver. The theory $\tilde{B}$ should have the same Higgs branch dimension
as the theory $\tilde{B}$. The graph mirror has the same Higgs branch as $\tilde{B}$.  Now we would like
to do some manipulation on the ``bad'' node so that the Higgs branch dimension is not changed. Let's consider
a $USp(2k)$ node, the Higgs branch contribution (include all the matter attached on it) is
\begin{equation}
N_f2k-(2k^2+k)={1\over 2}2k(2N_f-2k-1).
\end{equation}
For a bad quiver, we may wonder to replace its rank $2k$ with $2N_f-2k-1$ so the contribution
to the Higgs branch is the same. We can not do this since $2N_f-2k-1$ is a odd number.
We can try to replace $2k$ with
\begin{equation}
2k^{'}=2N_f-2k-2=2k+2(N_f-2k-1)=2k+2e,
\end{equation}
This is similar to the unitary case. However, after doing this, the Higgs branch contribution of this node
is increased by 2e.

For the $SO(N_c)$ gauge theory, one can do the similar calculation as above, the final result is that if
we replace the rank of the gauge group by $N_c^{'}=N_c+2e$. The Higgs branch contribution of this node
is increased by $-2e$.

Suppose we have a node with negative excess number $e$, after doing the manipulation, the Higgs branch
dimension is changed by $2e$ (we assume that this node is $USp$ type, $SO$ type is similar), assume one
of the adjacent node originally has excess number $e_1$, its excess number becomes $e_1^{'}=e_1+e$. First,
we should ensure that $e_1^{'}<0$, and we replace its rank follows our rule,  the Higgs
branch dimension is changed by $-2e_1^{'}$, to cancel the change of the $USp$ node, we have the relation
\begin{equation}
e_1=0.
\end{equation}
One may wonder we can do the manipulation on both adjacent nodes to cancel the contribution,
a little bit calculation shows that this is possible only in the case $e_1=e_2=0$.

Our conjecture is that we only do the manipulation on those ``bad'' nodes one of whose
adjacent node is ``balanced''. We continue this process until no nodes satisfy this condition.
This constraint makes sense, the ``bad'' node on the original quiver leg does not satisfy this condition so
we do not need to do the manipulation.

Let's also give a simple example to illustrate the idea. We take an theory $A$ for which we have a lagrangian
description (the general case are really the same). See fig.~\ref{mr11} for the details. The flavor symmetry on
the Coulomb branch of the quiver in fig.~\ref{mr11}(f) is changed from $SO(2)$ to $SO(4)$, which is exactly the
flavor symmetry on $USp(2)$ node in fig.~\ref{mr11}(a) (original, we have two half-hypermultiplets on $USp(2)$ node,
the flavor symmetry is $SO(2)$, after adding two more half-hypermultiplets, the flavor symmetry is changed
$USp(4)$).

With this construction, We can reproduce the results from \cite{Feng}.
In fact, we have constructed a large class of new mirror pairs for which the theory $\tilde{A}$ involves strongly
coupled matter.

An important application of this ``D5'' brane probe is that we can read the weakly coupled gauge group by
counting the change of the dimension of the Coulomb branch of the mirror. One subtly we should mention is
that for the $USp$ leg, we just add one ``D5'' brane and count the change of the Coulomb branch dimension in
the mirror. However, the $SO$ node is ``bad'', so we need to first add one ``D5'' brane to make it good, and
then add another "D5" brane to probe the rank of the gauge group. This is the only tool we know to completely
determine the generalized quiver from $D_N$ theory. One may also determines the matter contents as we
do for the unitary case.

One can also extend those consideration to the higher genus theory of the $D_N$ type.

\section{Gauging $U(1)$}
For all the theories considered in \cite{BYX} and this paper, the theory $A$ has $SU(k)$ gauge groups while
in the mirror $B$, there is no fundamentals attached on any quiver node. In this section, we will show how
the mirror changes if some of the $U(1)$ symmetry of theory $A$ is gauged.

The rule is quite simple, when there is a U(1) symmetry in the $A$ theory, there is a $U(1)$ gauge group in the mirror.
If we gauge the U(1) symmetry of $A$ to get theory $\tilde{A}$, the Higgs branch dimension of $\tilde{A}$ is decreased by 1
while the Coulomb branch dimension is increased by 1 comparing with $A$. To match this counting, we should ungauge the U(1) node of $B$ to get
a theory $\tilde{B}$ whose Higgs branch is increased by 1 while the Coulomb branch is decreased by 1 comparing with $B$ . This is
in agreement with the prediction of mirror symmetry.

The theory $\tilde{A}$ loses a mass parameter while $\tilde{B}$ loses a FI parameter; The Higgs branch of $\tilde{A}$
loses a $U(1)$ global symmetry while $\tilde{B}$ loses a $U(1)$ global symmetry in Coulomb branch. Those are also
in agreement with Mirror symmetry.

Consider the example in fig.~\ref{mr8}(c), when we gauge the U(1) symmetry, the $A$ theory is U(N) theory an adjoint and
two fundamentals, the $B$ theory is the quiver in the  right of fig.~\ref{mr8}(c) with the U(1) node uncaged, this is in agreement
with the result in \cite{Ooguri1}.

With this gauging trick, we can add some fundamentals on the central nodes (nodes with at
least three instrumentals attached on it)in the mirror. Notice that we can not add fundamentals on
the nodes on quiver tail attached to the central node.

\section{General quiver tail}
In the above generalizations, we do not change the boundary condition on the external leg of the graph, so
the quiver tail is the same. Our theory $A$ is simply a chain of simple unitary nodes coupled
with fundamentals (sometimes antisymmetric matter) and strongly coupled matter. Theory $A$ does not
have a lagrangian description in general. Our theory $B$ is always a standard quiver gauge theory.

In this section, we want to change the boundary condition so that we have general quiver tail
attached to the central node. In particular, we will allow quiver tails whose nodes can
have the fundamental hypermultiplets.

To proceed, we need to reinterpret the graph representation of the theory $A$ in terms of ${\cal N}=4$
SYM on a half space. The general boundary condition of ${\cal N}=4$ theory can be labeled by
$(\rho,H,B)$ \cite{GW1}, where $\rho$ is the homomorphism $\rho:su(2)\rightarrow g$, $H$ is the commutate of the $\rho$
in G, $B$ is a three dimensional theory which is coupled to $H$.

Consider a theory $A$ whose graph representation only has one internal leg and four external legs,
and we assume that the weakly coupled gauge group on the internal leg is $SU(N)$. This theory can be interpreted
as gauging two matter systems $T_1,T_2$ which are represented by trivalent graphs . As we proved in \cite{dan1}, these
two matter systems must have $SU(N)$ flavor symmetry, the gauge group is derived by gauging the diagonal $SU(N)$ symmetry.
These two matter systems are irreducible (they have
Coulomb branch parameters with dimension $N$), so their 3d mirrors are ``good'' quivers.
Consider one end of the internal leg, the boundary
condition is just $(0,SU(N),T_1)$, similar thing can be said at the other end.

The mirror theory $B$ can be found by first finding the mirror of the matter system $T_1, T_2$ and
then gluing them together. The mirror of the matter system is also the star-shaped quiver with a quiver tail which
is $T(SU(N))$ (see \cite{GW2} for its definition) corresponding to the $SU(N)$ flavor symmetry. The result
of gluing is just annihilate those two $T(SU(N))$ tails
and we are left with only one central node, we refer this as the gluing process to get the mirror $B$,
which is a counterpart of gauging for the original theory $A$. See figure 8 in \cite{BYX}.

Let's count the Coulomb branch and Higgs dimension of the theory  from gluing $T_1$ and $T_2$. For theory $A$, the Coulomb branch dimension is the
sum of the Coulomb branch of three parts $(C_1+C_2+(N-1))$, where $C_1$ and $C_2$ is the Coulomb branch dimension of $T_1$ and $T_2$.
The Higgs branch dimension is the  $(H_1+H_2-(N^2-1))$, here $H_1$ and $H_2$ is the Higgs branch dimension of $T_1,T_2$.
In the mirror, before the annihilation, the Coulomb branch dimension is simply $H_1+H_2$ by mirror symmetry.
After annihilating two $T(SU(N))$ legs, the Coulomb branch dimension is decreased by $N^2-1$;
and the Higgs branch dimension is increased by $N-1$, this agrees with theory $A$ using mirror symmetry.
\begin{figure}
\begin{center}
\includegraphics[width=4in]{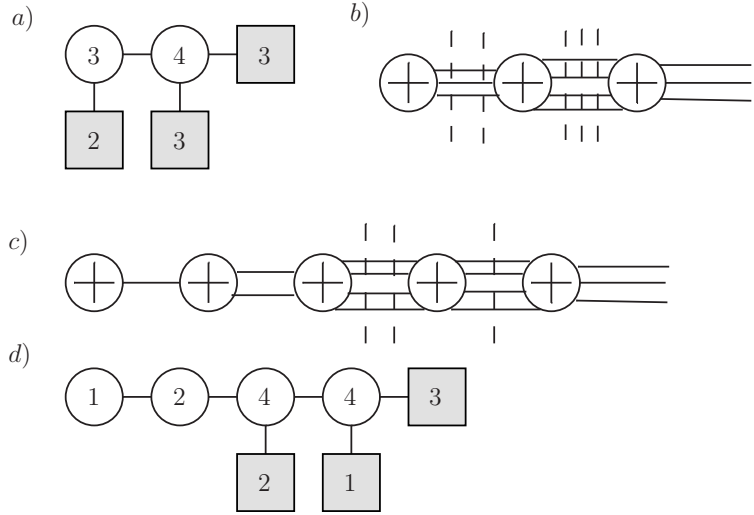}
\end{center}
\caption{(a)A linear quiver which we called theory $B$ in the main text, it has a $SU(3)$ symmetry which is
used to coupled to the $T_(SU(N))$. (b)The brane construction for the quiver in (a), here crosses represent
NS5 branes, vertical dash lines represent the D5 brane, horizonal lines are D3 branes. (c) The S-dual
brane configuration of (b), we have done a brane rearrangement so there is a gauge theory interpretation.
(d) The quiver representation of (c), which is the theory $B^{\vee}$ in the main text.
}
\label{mr12}
\end{figure}

\begin{figure}
\begin{center}
\includegraphics[width=4in]{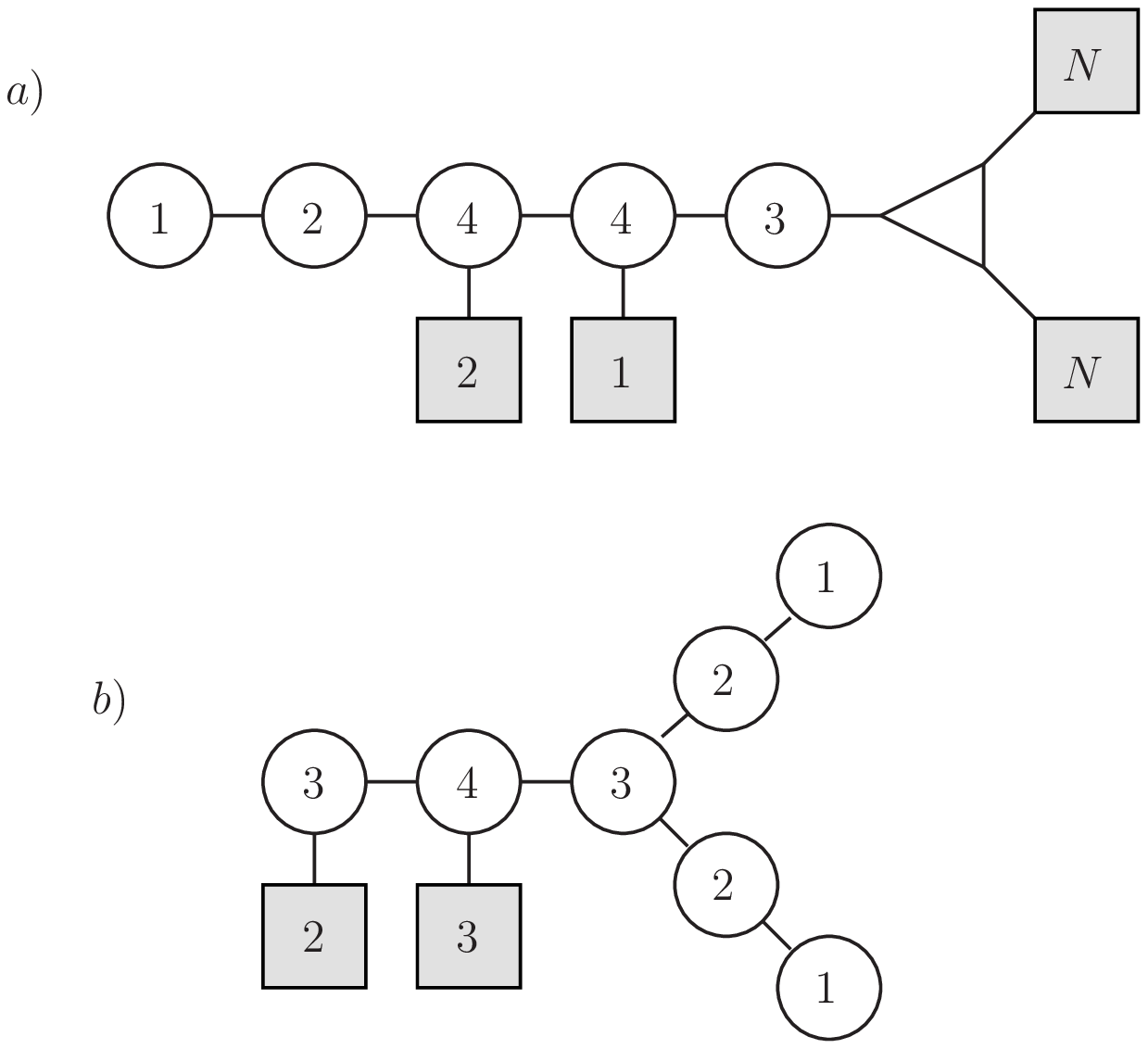}
\end{center}
\caption{(a)A quiver with $T_3$ factor on one end of $SU(3)$ gauge group, on the other
end, it is coupled to the theory $B^{\vee}$ as depicted in fig. 12(d). (b) The mirror to theory (a),
we use the quiver $B$ in Figure 12(a).
}
\label{mr13}
\end{figure}

Now we can replace the theory $T_1$ and $T_2$ by any other ``good'' theories $B^{\vee}$ with $SU(N)$ flavor symmetry
to form a theory $\tilde{A}$. As long as we know the mirror of $B^{\vee}$, we may find the mirror theory $\tilde{B}$. If
the mirror of $T$ has a quiver tail $T(SU(N))$, nothing will prevent us to find the mirror $\tilde{B}$ by simply
annihilating the $T(SU(N))$. Interestingly, in \cite{GW2}, a large class of those theories has been found.

Let's give a short review of their results. Suppose we have a boundary condition $(0, SU(N), B)$,
assume the dual boundary condition has full gauge symmetry.
the dual boundary condition is $(0,SU(N),B^{\vee})$ where $B^{\vee}$ is a SCFT living in the boundary. The mirror $\tilde{B}^{\vee}$ is
\begin{equation}\label{qv}
\tilde{B}^{\vee}=B_{SU(N)\times SU(N)}T(SU(N)).
\end{equation}
We assume that this is a ``good'' quiver, so the $SU(N)$ symmetry on the Coulomb branch of $T(SU(N))$ is the
$SU(N)$ symmetry of the Higgs branch of $B^{\vee}$.

Now we replace one of theory $T_1$ with theory $B^{\vee}$, the mirror is simply to replace two quiver tails from $T_1$
by $B$. The above consideration is quite general. Here we consider the case where the theory $B$ is a linear quiver,
since we require the quiver in \ref{qv} to be good quiver, the first node of this quiver should have rank $N_c\geq N+1$. B has
a $NS5-D5-D3$ brane constructions, we can find theory $B^{\vee}$ by doing S-duality on the brane systems. To have
a gauge theory interpretation, we may need to rearrange the branes, see \cite{HW,GW2} for more details. See fig.~\ref{mr12} for an example.

Now let's consider a SU(3) theory coupled with two $T_3$ theory, we replace one of $T_3$ with the theory $B^{\vee}$, which
we call the theory $\tilde{A}$, the mirror theory $\tilde{B}$ can be found from our general recipe. See fig.~\ref{mr13} for an example.

The theory $T_N$ can play an interesting role, for each $SU(N)$ flavor symmetry, we can attach a general quiver tail.
then the mirror is a star-shaped quiver with three general quiver tails. In fact, using $T_N$ theory, we can construct
the mirror theory with any number of general quiver tails.

The above consideration can be extended to the case even if the quiver \label{quiver} is ``bad'', the mirror is still given
by a star-shaped quiver as we described earlier, but the gauge group on the leg is not $SU(N)$ but broken down to a subgroup, we can use
the method in section 3 to determine the theory $A$.

\section{Irregular singularity}
The mirror theory we considered so far are almost linear quiver with only one bi-fundamental connecting
the quiver nodes. The shape of the quiver is quite simple. In this section, we will see more general type of
quivers.

There are other four dimensional ${\cal N}=2$ field theories constructed from six dimensional $(0,2)$ SCFT.
In the examples discussed in \cite{BYX}, one consider the compactification on Riemann surface with regular
singularities, this defines a four dimensional ${\cal N}=2$ SCFR \cite{dan3}. However, one can also
consider irregular singularities \cite{dan2, witten} on the Riemann surface. This defines a four dimensional
theory $A$ which we will study in detail elsewhere. In this section, we will study the mirror theory for some
of those theories which have already been mentioned in \cite{dan2}.

The moduli space of the Hitchin equation with irregular singularity
is the Coulomb branch of the four dimensional theory on a circle with radius $R$. In the deep IR limit, there
is a three dimensional ${\cal N}=4$ SCFT which  we call theory $A$.
We want to find the mirror theory $B$. In the case of regular singularities, we attach a quiver
tail to each of the singularity and then glue the common $SU(N)$ nodes together.

The procedure for the irregular singularity case is quite similar. We need to define a quiver to the irregular
singularity. After doing this, we  glue those quiver tails of the regular singularities to the quiver associated with
the irregular singularity. So the problem is reduced to find the quiver tail of the irregular singularity. The
general story of irregular singularity is quite complicated and we do not intend to consider the general case in this paper.
We only consider some simple cases and discuss the general theory elsewhere. Although as simple as the theory $A$ we
consider in this paper, the mirror seems to have some new features: more than one bi-fundamentals and exotic quiver shape.
We should mention that, in mathematics literature, the moduli space of Hitchin's equation in complex structure J
has been given a quiver approximation \cite{Bol2}, in physics language, we extend this observation to the
level of Hyperkahler structure, moreover, the Coulomb branch of the quiver is identified with the Higgs
branch of the theory $A$, which is not recognized in mathematics literature.

\subsection{$A_1$ Theory}
The $SU(2)$ Hitchin system defined on Riemann surface (in this paper, we only consider Riemann sphere)
has only two types of irregular singularities \cite{witten,dan2}, we write
the form of the holomorphic Higgs field
\begin{eqnarray}
\Phi={A\over z^n}+....\nonumber\\
\Phi={A\over z^{n-1/2}}+....,
\end{eqnarray}
where $A$ is a diagonal matrix. we call them Type I and Type II singularity respectively. When we put such a singularity on the Riemann
surface, we get a four dimensional ${\cal N}=2$ theory $A$. One can add other regular singularities on the Riemann sphere. The
moduli space of Hitchin's equation in complex structure $I$ has the famous Hitchin's fibration, which is identified
with the Seiberg-Witten fibration of the four dimensional theory. In particular, the Coulomb branch dimension of
four dimensional theory has half the dimension of the Hitchin's moduli space. The spectral curve of the Hitchin system is
\begin{equation}
det(x-\Phi)=0, \label{spectral}
\end{equation}
which is just the Seiberg-Witten curve. The total dimension of Hitchin's moduli space is equal to the contribution of each singularity
and minus the global contribution $2(dim{G}-r)=6$, here $G=SU(2)$, $r$ is the rank of the gauge group. The local contribution of regular singularity is $2$ \cite{dan2}, and $2n$ for the irregular singularity. We only consider just one irregular singularity on this paper(when
there are more than one irregular singularities, the mirror is not a quiver).

The mass parameter for the four dimensional theory is encoded in the residue of the Higgs field. So for type I singularity,
there is one mass parameter. However, there is no mass parameter for type II singularity since the residue term is
not allowed because of monodromy. We want to attach a quiver with these irregular singularities. In the case of regular
singularity, the dimension of the Higgs branch of the quiver tail only accounts for the local dimension of the singularity.
In the case of irregular singularity, we should include the global contribution to the irregular singularity: the Higgs
branch of the quiver should have dimension $n-3$. The quiver should have one $U(1)$ factor for Type I node and no $U(1)$ factor for
type II node, since the FI term for the $U(1)$ would correspond to the mass parameter (these are not true in general, since
we have exotic IR behavior for some theories, in these cases, one can not tell what exactly happens in the IR from the UV theory).

\begin{figure}
\begin{center}
\includegraphics[width=3in]{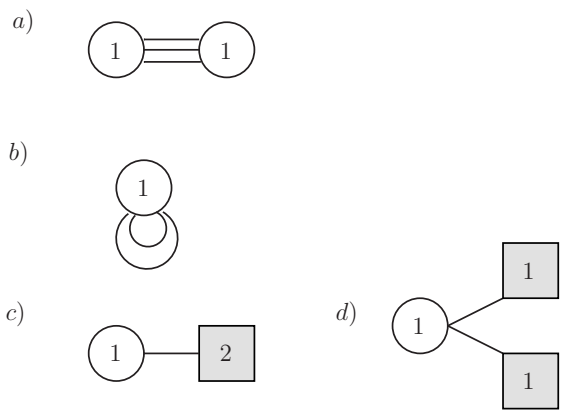}
\end{center}
\caption{(a) The quiver for type I irregular singularity, here $n=5$.
(b) The quiver for type II irregular singularity, here $n=5$, here we have adjoint matter on $U(1)$. (c) The quiver tail for a regular singularity.
(d) We spray the $SU(2)$ symmetry into two $U(1)$s.
}
\label{mr14}
\end{figure}

With these considerations, we have the following conjecture for the quiver attached on the irregular singularity:
for the Type I singularity, we have two nodes with U(1) group, and there are
$(n-2)$ bifundamentals connecting them, one of the $U(1)$ is decoupled, so we only have one FI term;
There are $n-3$ mass terms for the bi-fundamentals, this means that the original theory has $n-3$ ``hidden''  FI
terms. The origin of these "hidden" FI terms will be discussed elsewhere. One example is shown in fig.~\ref{mr14}(a).

For Type II singularity, if $n\geq 5$ we have only one $U(1)$ node with $n-2$ lines connect to itself, these are the adjoints for
the $U(1)$ which are just the fundamentals. This is exactly like the mirror for the high genus theory in the context of regular
singularities \cite{BYX}. There are also enhanced global symmetry in the mirror
and there are extra $n-3$ mass parameters which correspond to the ``hidden'' FI term in the original theory $A$. Follow the
analogy with the higher genus theory, $n=4$ corresponds to genus $1$ case. In the genus one case,
the massless limit has enhanced supersymmetry, and the mirror has only one adjoints while the massive limit has an extra $U(1)$
node. For the irregular singularity here, we conjecture that the mirror corresponds to the massless limit of the genus $1$ case, so we only have one adjoint, the mirror is indeed $U(1)$ with one fundamental, we will confirm this later. Similarly, for $n=3$, there is
only one $U(1)$ node, in this case, there is no meaning to consider this irregular singularity alone, this is only a recipe
to form the mirror quiver when there are extra regular singularities. For this class of singularities, we show one example in fig.~\ref{mr14}(b).
\begin{figure}
\begin{center}
\includegraphics[width=3in]{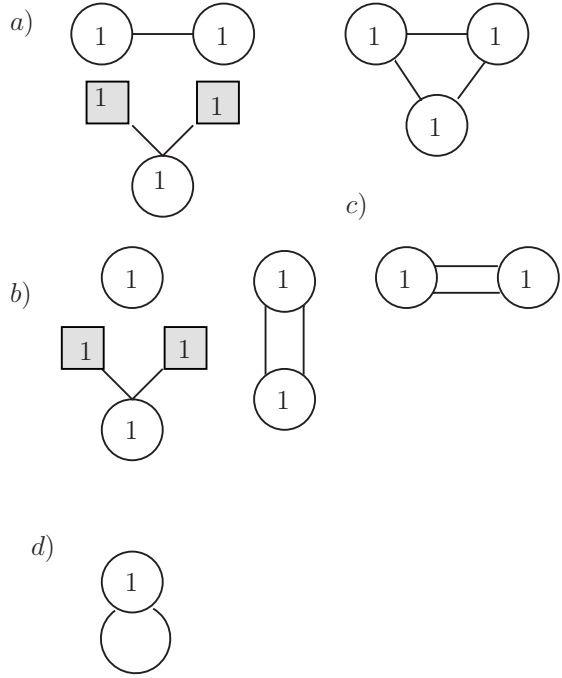}
\end{center}
\caption{(a)The mirror for the $A_2$ Argyres-Douglas point, there is one Type I irregular singularity with $n=3$ and
one regular singularity, we glue them together to form a quiver on the right. (b) For $A_1$ Argyres-Douglas point,
there is a type II irregular singularity with $n=3$ and a regular singularity, after gluing them, we get
a $A_1$ affine dynkin diagram. (c) Another representation of $A_1$ Argyres-Douglas point, only one Type I irregular
singularity with $n=4$ needed, the resulting quiver is the same as representation $(b)$. (d) For $A_0$ Argyres-Douglas
point, only one Type II singularity with $n=4$ is needed, the resulting mirror is just the $A_0$ affine dynkin diagram.
}
\label{mr15}
\end{figure}

When there are other regular singularities, the quiver tail is shown in fig.~\ref{mr14}(c). We spray the node as in fig.~\ref{mr14}(d).
To connect the regular singularity to irregular singularity, we just gauge the $U(1)$ node: In type I case, they are
gauged separately; in type II case, they are gauged on the same node. Since the Higgs branch of this quiver tail
accounts the local dimension of the regular singularity to the Hitchin's moduli space even after spraying, the Higgs branch of
the whole quiver is the same as the Coulomb branch of the theory $A$.

We can consider some examples studied in \cite{dan2}. In that paper, we claim that the Argyres-Douglas $A_2$ theory
is derived with one Type I irregular singularity with $n=3$ and a regular singularity. The Seiberg-Witten curve as from \ref{spectral} is
\begin{equation}
x^2=z^2+u_1z+u_2+{u_3\over z}+{m^2\over z^2}.
\end{equation}
We put the irregular singularity at $z=\infty$ and regular singularity at $z=0$. Let's count the scaling
dimension of the operators in the curve. The Seiberg-Witten differential is $\lambda=xdz$; We require
its dimension to be $1$, together with the form of the Seiberg-Witten curve, we have the scaling dimension of
$x$ and $z$:
\begin{equation}
[x]={1\over 2},[z]={1\over2}.
\end{equation}
One can easily get the scaling dimension of the operators: $[u_1]={1\over 2},[u_2]=1,[u_3]={3\over 2},[m]=1$,
which is the same as the $A_2$ Argyres-Douglas (AD) points as shown in \cite{AD}. The Coulomb branch dimension of this theory is $1$,
and the Higgs branch dimension is $2$. The flavor symmetry is $SU(3)$ which is not easy to see from our six dimensional
description, since there is only manifest $SU(2)\times U(1)$ flavor symmetry. We will see below that the
we can see the enhanced symmetry from the mirror.

The mirror quiver is depicted in fig.~\ref{mr15}(a), we show how to glue the quiver from the regular
singularity and the irregular singularity together. Let's check that it gives the correct mirror description: The Higgs
branch dimension is 1 and Coulomb branch is $2$, which agrees with the prediction from mirror symmetry.
Since the mirror quiver has a chain of balanced quiver
with two nodes (one $U(1)$ is decoupled), the symmetry in the Coulomb branch is $SU(3)$ which is exactly
the flavor symmetry of the $A_2$ theory.

For $A_1$ AD points, there are two construction: one Type II irregular singularity with $n=3$ and a regular
singularity; or we can have just one Type I irregular singularity with $n=4$. For the first representation,
the spectral curve is
\begin{equation}
x^2=z+u_1+{u_2\over z}+{m^2\over z^2}.
\end{equation}
We put irregular singularity at $z=\infty$ and regular singularity at $z=0$. The scaling dimension is $[x]={1\over 3},[z]={2\over 3}$.
The scaling dimension of the spectrum is $[u_1]={2\over 3},[u_2]={4\over 3},[m]=1$, which is the same as  the $A_1$ points \cite{AD}.

For another representation, the spectral curve is
\begin{equation}
x^2=z^4+u_1z^2+mz^3+u_2.
\end{equation}
We have shift the origin so $z$ term is absent. The spectrum is $[u_1]={2\over3},[m]=1,[u_2]={4\over3}$, which is
the same as the above representation.

The $A_1$ AD point has Coulomb branch dimension $1$ and Higgs branch dimension $1$, the flavor symmetry is $SU(2)$.
The 3d mirror $B$ are the same as we can see in fig.~\ref{mr15}(b), this also justifies that
these two descriptions are equivalent. The mirror has Coulomb branch dimension $1$ and Higgs branch dimension $1$,
the symmetry on the Coulomb branch is $SU(2)$, these are all in agreement with the mirror symmetry.

$A_0$ theory is defined on a sphere with a Type II irregular singularity with $n=4$. The spectral curve is
\begin{equation}
x^2=z^3+u_1z+u_2.
\end{equation}
The spectrum is $[u_1]={4\over 5},[u_2]={6\over5}$, which is the same as shown in \cite{AD}. The Coulomb
branch of $A_0$ theory is $1$ and the Higgs branch is $0$.

The mirror for $A_0$ theory is shown in fig.~\ref{mr15}(c), which is just $U(1)$ with one fundamentals. In the
deep IR, it is just a twisted free hypermultiplet, so the Coulomb branch is 1, and the Higgs  branch is $0$ (see e.g. \cite{GW2}).
In this case, the mirror symmetry does not match the Coulomb branch to Higgs branch, but match the Coulomb branch
to Coulomb branch, this kind of phenomenon has also been observed in \cite{IR}

\begin{figure}
\begin{center}
\includegraphics[width=2.5in]{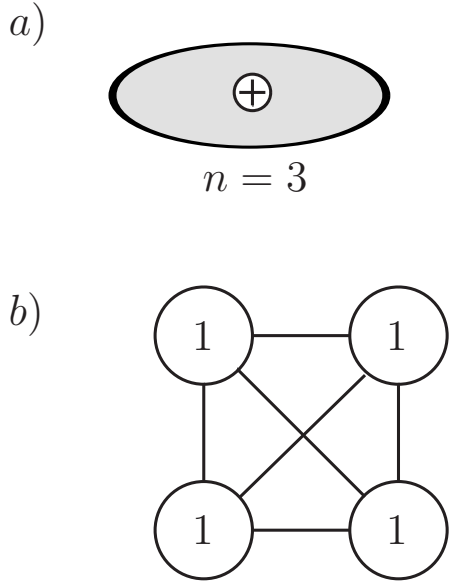}
\end{center}
\caption{(a) A four dimensional superconformal field theory derived from six dimensional $(0,2)$ theory on a
sphere with one irregular singularity. (b) The mirror to theory (a).
}
\label{mr16}
\end{figure}

Notice that the mirror are just the affine dynkin diagram of the corresponding type for the Argyres-Douglas points.
The singular fibre is classified by Kodaira with type $A_0, A_1, A_2, D_4, E_6, E_7, E_8$. Four dimensional
superconformal field theory with these curves are found \cite{Doug,AD,min1,min2}. The mirror theory of IR limit of the
three dimensional cousin are just given by the corresponding ADE affine dynkin diagram. Interestingly, the
Coulomb branch of these isolated SCFT in three dimensions are the ALE space of the corresponding type.

\subsection{$A_{N-1}$ Theory}

The classification of the irregular singularity for rank N theory is quite complicated. Here we only consider
the most simple irregular singularity.
\begin{equation}
\Phi={A\over z^n}+...
\end{equation}
where $A$ is the diagonal matrix with distinct eigenvalues. The Hitchin equation with this kind of
singularity has been considered in detail in the gauge theory approach to  Geometric Langlands program \cite{witten}.
The local dimension of just one singularity is
\begin{equation}
n(dim{G}-r)
\end{equation}
where $r$ is the rank of the gauge group and $G$ is $SU(N)$ in the present context. The total dimension of the Hitchin's
moduli space with just one such irregular singularity is
\begin{equation}
n(N^2-N)-2(N^2-1).
\end{equation}
So the four dimensional theory $A$ has Coulomb branch dimension
\begin{equation}
d_C={1\over 2}n(N^2-N)-(N^2-1)={1\over2}(n-2)(N^2-N)-(N-1).
\end{equation}
There are a total of $N-1$ mass parameters.

When we compactify six dimensional theory on a sphere with such a singularity,  we get a four dimensional SCFT $A$. The
spectrum can be worked out similar as the AD points, we will give it elsewhere, here we  consider its mirror $B$.
The mirror to this theory is quite simple, there are $N$ nodes with $U(1)$ gauge groups and there are $(n-2)$ bi-fundamentals
connecting each pair of nodes. The Higgs branch of this quiver $B$ is
\begin{equation}
(n-2){1\over 2}(N^2-N)-(N-1),
\end{equation}
which is exactly the Coulomb branch dimension of $A$.
See fig.~\ref{mr16} for an example.

\section{Conclusion}
In this work, we systematically study the interesting mirror symmetry for three dimensional
${\cal N}=4$ theory. With these construction, we reproduce most of the mirror pairs considered
before and found a large class of new family of mirror pairs. The addition of more fundamentals
can be used to probe the four dimensional generalized quiver gauge theory, in fact, we can
completely determine the weakly coupled gauge group and the matter contents using
3d mirror symmetry.

In $D_N$ case, one need to consider outer automorphism to describe conventional orthosymplectic
quiver. This automorphisms are useful to understand four dimensional S dualities \cite{vafa2,argy2,yu}.
$A_{2N-1}$ theory also has the $z_2$ outer automorphism; We can construct
four dimensional theory with these automorphism and ask what is its 3d mirror. The mirror
should have the quiver tail with unitary groups and orthosymplectic groups.
It is interesting
to work our the details.

In this paper, we discuss the mirror symmetry in the sense of the IR limit of three dimensional
theory. The mirror symmetry can be indeed extended to the finite gauge couplings, but in this case,
the mirror theory is indeed macroscopically four dimensional, see the discussion in \cite{HW}.
These extended mirror symmetry has interesting relation with the Hyperkahler manifold \cite{dan5}.

Another generalization is to consider adding Chern-Simons terms and consider how the mirror
changes, See \cite{Kapustin1} in the abelian case. It would be interesting to work out
the non-abelian Chern-Simons term. It is also interesting to extend to ${\cal N}=2$ supersymmetric
theory \cite{2susy}.

The irregular singularities are very interesting, it is inevitable if we want to use Hitchin's
equation to describe the asymptotical free theories. In $A_1$ case, the irregular singularities
are completely classified and the Argyres-Douglas theory can be described in this way, we
work out its three dimensional mirror in this paper. We only consider some simple
examples, it is interesting to work out the general case. We hope to return to this in the
near future.

Three dimensional mirror symmetry has been checked using the localization method \cite{Kapustin3}, it
is interesting to calculate the partition function of those star-shaped quiver, by this
we can learn more about four dimensional generalized quiver gauge theory.

\begin{flushleft}
\textbf{Acknowledgments}
\end{flushleft}
It is a pleasure to thank Francesco Bebini, Davide Gaiotto and Yuji Tachikawa
for helpful comments and useful discussions. This research was supported in part by the Mitchell-Heep chair in
High Energy Physics (CMC) and by the DOE grant DEFG03-95-Er-40917.


\begin{thebibliography}{999}
\bibitem{Gaiotto1} D. Gaiotto, ``N=2 dualities'', arXiv:0904.2715.
\bibitem{Argy}    P.C.Argyres and N.Seiberg, ``S-Duality in $N=2$
Supersymmeric Gauge Theories,'' JHEP 0712 (2007) 088
[arXiv:0711.0054][hep-th].
\bibitem{3d}N.Seiberg, E.Witten, ``Gauge Dynamics And Compactification To Three Dimensions,'' arXiv:hep-th/9607163.
\bibitem{Mr}K. A. Intriligator and N. Seiberg, ``Mirror Symmetry in Three Dimensional Gauge
Theories,'' Phys. Lett. B387 (1996) 513每519, arXiv:hep-th/9607207.
\bibitem{BYX} F. Benini, Y. Tachikawa, D. Xie, ``Mirrors of 3d Sicilian Theories,'' JHEP 1009:063,2010.
[arXiv:1007.0992].
\bibitem{GW1}D.Gaiotto, E.Witten, ``Supersymmetric Boundary Conditions in N=4 Super Yang-Mills Theory,'' arXiv:0804.2902.
\bibitem{GW2}D.Gaiotto, E.Witten, ``S-Duality of Boundary Conditions In N=4 Super Yang-Mills Theory,'' arXiv:0807.3720.


\bibitem{Ooguri1} J. de Boer, K. Hori, H. Ooguri, and Y. Oz, ``Mirror Symmetry in Three-Dimensional
Gauge Theories, Quivers and D-Branes,'' Nucl. Phys. B493 (1997) 101每147,
arXiv:hep-th/9611063.
\bibitem{Ooguri2} J. de Boer, K. Hori, H. Ooguri, Y. Oz, and Z. Yin, ``Mirror Symmetry in
Three-Dimensional Gauge Theories, $SL(2,Z)$ and D-Brane Moduli Spaces,'' Nucl. Phys.
B493 (1997) 148每176, arXiv:hep-th/9612131.
\bibitem{Feng}B. Feng and A. Hanany, ``Mirror symmetry by O3-planes,'' JHEP 11 (2000) 033,
arXiv:hep-th/0004092.
\bibitem{HW}A. Hanany and E. Witten, ※Type IIB Superstrings, BPS Monopoles, and
Three-Dimensional Gauge Dynamics,'' Nucl. Phys. B492 (1997) 152每190,
arXiv:hep-th/9611230.
\bibitem{Kapustin1} A. Kapustin and M. J. Strassler, ``On Mirror Symmetry in Three Dimensional Abelian
Gauge Theories,'' JHEP 04 (1999) 021, arXiv:hep-th/9902033.
\bibitem{Kapustin2}A. Kapustin, ``$D_n$ Quivers From Branes,'' JHEP 9812:015,1998, arXiv:hep-th/9806238.
\bibitem{Vafa}K. Hori, H. Ooguri, and C. Vafa, ``Non-Abelian Conifold Transitions and N=4 Dualities in Three Dimensions,'' Nucl.Phys.B504:147-174,1997, arXiv:hep-th/9705220

\bibitem{dan1}D. Nanopoulos and D. Xie,   ``N=2 Generalized Superconformal Quiver Gauge Theory,'' arXiv:1006.3486.
\bibitem{dan2}D. Nanopoulos and D. Xie,   ``Hitchin Equation, Irregular Singularity, and $N=2$ Asymptotical Free Theories,''
arXiv:1005.1350.
\bibitem{Mon}V.Borokhov, A. Kapustin, X.K. Wu, ``Monopole Operators and Mirror Symmetry in Three Dimensions,''
	JHEP 0212 (2002) 044, arXiv:hep-th/0207074.
\bibitem{dan3}D. Nanopoulos and D. Xie, ``Hitchin Equation, Singularity, and N=2 Superconformal Field Theories,''
JHEP 1003:043,2010, arXiv:0911.1990.
\bibitem{Bol}P. Boalch, ``Quivers and difference Painlev∩e equations,'' in Groups and symmetries,
J. Harnard and P. Winternitz, eds., vol. 47 of CRM Proceedings and Lecture Notes.
AMS, 2009. arXiv:0706.2634 [math.AG].


\bibitem{SO} Y. Tachikawa, ``Six-dimensional DN theory and four-dimensional SO-USp quivers,'' JHEP 07
(2009) 067, [arXiv:0905.4074].
\bibitem{web} F. Benini, S. Benvenuti, and Y. Tachikawa, ``Webs of five-branes and $N=2$ superconformal field
theories,'' JHEP 09 (2009) 052, [arXiv:0906.0359].
\bibitem{N1}K. Maruyoshi, M. Taki, S. Terashima, and F. Yagi, ``New Seiberg Dualities from $N=2$ Dualities,''
JHEP 09 (2009) 086, [arXiv:0907.2625].
\bibitem{dan4}D. Nanopoulos and D. Xie, ``$N=2$ SU Quiver with USP Ends or SU Ends with Antisymmetric
Matter,'' JHEP 08 (2009) 108, [arXiv:0907.1651].
\bibitem{yuji2}F. Benini, Y. Tachikawa, and B. Wecht, ``Sicilian Gauge theories and N=1 Sualities,''
arXiv:0909.1327.
\bibitem{stony}A. Gadde, E. Pomoni, L. Rastelli, and S. S. Razamat, ``S-Duality and 2D Topological
QFT,'' JHEP 03 (2010) 032, arXiv:0910.2225 [hep-th].
\bibitem{stony1}A. Gadde, L. Rastelli, S. S. Razamat, and W. Yan, ``The Superconformal Index of the
E6 SCFT,'' arXiv:1003.4244 [hep-th].
\bibitem{moore} D.Gaiotto, G.W.Moore, A.Neitzke, ``Wall-crossing, Hitchin Systems, and the WKB Approximation,''
arXiv:0907.3987.
\bibitem{dist}O. Chacaltana, J. Distler, ``Tinkertoys for Gaiotto Duality,'' arXiv:1008.5203.
\bibitem{mal}D. Gaiotto and J. Maldacena, ``The Gravity Duals of N = 2 Superconformal Field
Theories,'' arXiv:0904.4466 [hep-th].
\bibitem{witten}E.Witten, ``Gauge Theory And Wild Ramification,'' [arXiv:0710.0631].
\bibitem{Bol2}P.Boalch, ``Irregular connections and Kac-Moody root systems,''
[arXiv:0806.1050][math].

\bibitem{Doug}P. C. Argyres, M. R. Douglas, ``New Phenomena in SU(3) Supersymmetric Gauge Theory,'' 	Nucl.Phys. B448 (1995) 93-126,
arXiv:hep-th/9505062.
\bibitem{AD}P.C. Argyres, M.R. Plesser, N. Seiberg, E. Witten, ``New N=2 Superconformal Field Theories in Four Dimensions,''
	Nucl.Phys.B461:71-84,1996, arXiv:hep-th/9511154.
\bibitem{IR}M. Berkooz, A. Kapustin, ``New IR Dualities in Supersymmetric Gauge Theory in Three Dimensions,''	JHEP 9902 (1999) 009,
arXiv:hep-th/9810257.
\bibitem{min1}J. A. Minahan and D. Nemeschansky, ``An $N = 2$ Superconformal Fixed Point with E6
Global Symmetry,'' Nucl. Phys. B482 (1996) 142每152, arXiv:hep-th/9608047.
\bibitem{min2}J. A. Minahan and D. Nemeschansky, ``Superconformal Fixed Points with EN Global
Symmetry,'' Nucl. Phys. B489 (1997) 24每46, arXiv:hep-th/9610076.
\bibitem{vafa2}C. Vafa, ``Geometric Origin of Montonen-Olive Duality,'' Adv. Theor. Math. Phys. 1
(1998) 158每166, arXiv:hep-th/9707131.
\bibitem{argy2}P. C. Argyres, J.R. Wittig,``Infinite Coupling Duals of $N=2$ Gauge Theories and New Rank 1 Superconformal Field Theories,''
	JHEP 0801:074,2008,arXiv:0712.2028.
\bibitem{yu}Y. Tachikawa, ``N=2 S-duality via Outer-automorphism Twists,'' arXiv:1009.0339.
\bibitem{dan5}D.Nanopoulous, D.Xie, work in progress.
\bibitem{2susy}O. Aharony, A. Hanany, K. Intriligator, N. Seiberg, M.J. Strassler, ``Aspects of N=2 Supersymmetric Gauge Theories in Three Dimensions,'' Nucl.Phys.B499:67-99,1997, arXiv:hep-th/9703110.
\bibitem{Kapustin3}A. Kapustin, B. Willett, I. Yaakov, ``Nonperturbative Tests of Three-Dimensional Dualities,'' arXiv:1003.5694.

\end{thebibliography}
\end{document}